\documentclass[reprint,superscriptaddress,amsmath,amssymb,aps,pra,showkeys,onecolumn,10pt]{revtex4-2}
\usepackage{graphicx}
\usepackage{dcolumn}
\usepackage{bm}
\usepackage[labelformat=empty, position=top, skip=1pt]{subcaption}
\usepackage{cases}
\usepackage[hidelinks]{hyperref}
\usepackage{orcidlink}
\usepackage{braket}
\usepackage{booktabs}
\usepackage{amsmath}
\usepackage{algorithm}
\usepackage{algorithmic}
\usepackage{float}
\usepackage{tikz}
\usepackage{comment}
\usetikzlibrary{quantikz2,shapes,arrows,calc,fit,backgrounds,shapes.multipart,patterns, patterns.meta, pgfplots.groupplots,shapes.arrows,positioning,arrows, chains, quotes,babel, math, arrows.meta, external, positioning, 3d, shadows, decorations.pathreplacing,calligraphy}
\tikzset{
	contour/.style={draw, thick, color=black},
	flow domain/.style={draw, thick, color=black, fill=black!10!white},
	void domain/.style={draw, thick, color=black, fill=white},
	fictitious domain/.style={draw=none, pattern={north east lines}, pattern color=black!50!white},
	sketch2D/.style={scale=1.0},
}

\usepackage{pgfplots}
\usepgfplotslibrary{colorbrewer,external}
\pgfplotsset{
    compat=newest,
    width=0.7\linewidth,
    height=0.4\linewidth,
    xticklabel style = {font=\scriptsize},
    yticklabel style = {font=\scriptsize},
    scale only axis,
    xlabel=$x$,
    ylabel=$u$,
    yticklabel style={text width=2em},
    grid=major,
	cycle list/Set1-9,
	cycle multiindex* list={
		Set1-9\nextlist
		mark list*\nextlist
		solid, every maker/.append style={solid, draw=.!80!black, fill=.}\nextlist
	},
    every axis/.append style={line width = 1.3pt},
  colormap={inferno}{
     rgb(0)=(0.001462, 0.000466, 0.013866),
     rgb(15)=(0.037668, 0.025921, 0.132232),
     rgb(30)=(0.116656, 0.047574, 0.272321),
     rgb(45)=(0.217949, 0.036615, 0.383522),
     rgb(60)=(0.316282, 0.053490, 0.425116),
     rgb(75)=(0.410113, 0.087896, 0.433098),
     rgb(90)=(0.503493, 0.121575, 0.423356),
     rgb(105)=(0.596940, 0.154848, 0.398125),
     rgb(120)=(0.688653, 0.192239, 0.357603),
     rgb(135)=(0.775059, 0.239667, 0.303526),
     rgb(150)=(0.851384, 0.302260, 0.239636),
     rgb(165)=(0.912966, 0.381636, 0.169755),
     rgb(180)=(0.956852, 0.475356, 0.094695),
     rgb(195)=(0.981895, 0.579392, 0.026250),
     rgb(210)=(0.987464, 0.690366, 0.079990),
     rgb(225)=(0.973088, 0.805409, 0.216877),
     rgb(240)=(0.947594, 0.917399, 0.410665),
     rgb(255)=(0.988362, 0.998364, 0.644924),
     },
}
\pgfplotscreateplotcyclelist{gray}{ % define a new cycle list
 {gray, fill=gray!10,mark=-},
 {gray, fill=gray!20,mark=|},
 {gray, fill=gray!30,mark=o},
 {gray, fill=gray!40,mark=asterisk},
 {gray, fill=gray!50,mark=triangle},
 {gray, fill=gray!60,mark=diamond},
 {gray, fill=gray!70,mark=square},
 {gray, fill=gray!80,mark=x},
}

\definecolor{red}{RGB}{228,26,28}
\definecolor{blue}{RGB}{55,126,184}
\definecolor{green}{RGB}{77,175,74}
\definecolor{purple}{RGB}{152,78,163}
\definecolor{orange}{RGB}{255,127,0}
\definecolor{yellow}{RGB}{255,255,51}
\definecolor{brown}{RGB}{166,86,40}
\definecolor{pink}{RGB}{247,129,191}
\definecolor{gray}{RGB}{153,153,153}

\graphicspath{{figures/}}

\begin{document}

\title{Operator learning for efficient quantum computation}

\author{\href{mailto:paul.over@tuhh.de}{Paul Over}\orcidlink{0000-0001-7436-5254}}
\email[]{paul.over@tuhh.de}
\affiliation{
\href{https://tuhh.de/fds}{Institute for Fluid Dynamics and Ship Theory}, \href{https://ror.org/04bs1pb34}{Hamburg University of Technology}, Am Schwarzenberg-Campus 4, D-21073 Hamburg, Germany,}

\author{\href{mailto:sergio.bengoechea@tuhh.de}{Sergio Bengoechea}\orcidlink{0009-0001-8205-5878}}
\affiliation{
\href{https://tuhh.de/fds}{Institute for Fluid Dynamics and Ship Theory}, \href{https://ror.org/04bs1pb34}{Hamburg University of Technology}, Am Schwarzenberg-Campus 4, D-21073 Hamburg, Germany,}

\author{\href{mailto:leonardo.borello@planqc.eu}{Leonardo Borello Busilacchi}\orcidlink{0009-0009-7755-382X}}
\affiliation{
\href{https://planqc.eu}{PlanQC GmbH}, Münchener Str. 34, D-85748 Garching, Germany.
}

\author{\href{mailto:martin@planqc.eu}{Martin Kiffner}\orcidlink{0000-0002-8321-6768}}
\affiliation{
\href{https://planqc.eu}{PlanQC GmbH}, Münchener Str. 34, D-85748 Garching, Germany.
}
\author{\href{mailto:thomas.rung@tuhh.de}{Thomas Rung}\orcidlink{0000-0002-3454-1804}}
\affiliation{
\href{https://tuhh.de/fds}{Institute for Fluid Dynamics and Ship Theory}, \href{https://ror.org/04bs1pb34}{Hamburg University of Technology}, Am Schwarzenberg-Campus 4, D-21073 Hamburg, Germany,}

\author{\href{mailto:alexios.michailidis@planqc.eu}{Alexios A. Michailidis}\orcidlink{0000-0002-8443-1064}}
\email[]{alexios.michailidis@planqc.eu}
\affiliation{
\href{https://planqc.eu}{PlanQC GmbH}, Münchener Str. 34, D-85748 Garching, Germany.
}

\date{\today}

\begin{abstract}
An efficient implementation of quantum algorithms is often hindered by the lack of efficient primitives for operators and state preparation. 
This limits both the ability of near-term quantum hardware to simulate complex problems and the potential of fault-tolerant algorithms to achieve practical quantum advantage. 
To address this, we propose a framework that maps discrete operators onto compact, structured quantum circuits.
The resulting variational circuits can be tailored to the connectivity and long-range interaction of the target hardware. 
The learning process employs backpropagation together with a cost function that efficiently optimizes unitary operators and non-unitary --dense or sparse-- operators using only a single ancilla qubit for block encoding.
Additionally, we introduce a regularization term that reduces the approximation error.
The approach is validated for both quantum mechanical and engineering applications. 
In the former case, we learn propagators that arise in native quantum problems --such as quantum simulation and quantum chemistry-- and achieve improved resource scaling in comparison to standard Suzuki--Trotter expansions.
In the latter case, we demonstrate the approach's ability to implement the second-order central finite difference approximation of the Laplace operator --relevant for solving partial differential equations-- while improving upon current error metrics.
The final example deals with learning a dense, non-unitary operator that arises in the analysis of inviscid potential flow around an airfoil. 
These results motivate the extension of the framework to a broader class of operator-synthesis problems beyond the prototypical engineering and quantum applications considered here.
\end{abstract}


\maketitle

%%%%%%%%%%%%%%%%%%%%%%%%%%%%%%%%%%%%%%%%%%%%%%%%%%%%%%%%%%%%%%%%%%%%
%%  INTRO
%%%%%%%%%%%%%%%%%%%%%%%%%%%%%%%%%%%%%%%%%%%%%%%%%%%%%%%%%%%%%%%%%%%%
\section{Introduction}
\label{sec:intro}
Over the past decades, numerical methods for scientific computing have achieved significant advances~\cite{Press2007}. 
Nevertheless, in-depth analysis requires extensive high-performance computing resources, as classical algorithms exhibit inherent scaling limitations as the system size increases~\cite{Nasa2030}. 
At the same time, the challenge of further reducing transistor sizes~\cite{Khan2018,Esmaeilzadeh2011} restricts performance improvements in classical hardware.
In this context, Quantum Processing Units (QPUs) offer a transformative alternative by leveraging the exponentially large vector spaces present in many-body quantum mechanics. 
QPUs enable novel computational paradigms in which quantum algorithms might achieve substantial improvements in runtime and memory complexity, while potentially offering superior energy efficiency compared to classical architectures~\cite{Givi2020}.

%%Universal Quantum Computer 
Despite the great prospects of universal QPUs~\cite{Bluvstein2025,Campbell2017}, current hardware limitations prevent the observation of practical quantum advantage. 
Today's Noisy Intermediate-Scale Quantum (NISQ) devices suffer from finite coherence times, gate errors, and restricted qubit connectivity, which constrain achievable circuit depth and significantly increase the cost of multi-qubit operations~\cite{Arute2019,Preskill2018,Camps2020}.
These hardware constraints limit the complexity of circuits that can be executed reliably, making compact and noise-aware circuit design essential for near-term applications.

Even looking beyond NISQ, early fault-tolerant devices face a distinct but equally serious challenge: a disparity in clock speeds between classical processors and logical quantum qubits, typically estimated to span several orders of magnitude~\cite{Hoefler2023}.
This means that quantum algorithms must deliver sufficient algorithmic speedup to compensate for this inherent slowdown, highlighting the importance of early-stage near-hardware algorithmic design.
In practice, asymptotic quantum speedups are frequently offset by large constant prefactors arising from block-encoding overheads, Quantum State Preparation (QSP) costs, and measurement complexity~\cite{Hoefler2023}. 
In the worst case, the decomposition of generic multi-qubit unitaries and QSP requires resources that scale exponentially with the number of qubits~\cite{Mottonen2005}, further eroding any theoretical advantage. 
Practical algorithms must therefore be carefully optimized and tailored to specific target platforms -- whether superconducting~\cite{Arute2019}, neutral atom~\cite{Evered2023}, trapped ion~\cite{Debnath2016}, or photonic~\cite{Wang2020} hardware -- highlighting the critical importance of early-stage algorithmic development and hardware-software co-design.

% % Classical to Quantum %%%%%%%%%%%%%%%
Quantum Circuit Synthesis (QCS) provides a unified framework for translating vectors and matrices into quantum gate sequences, encompassing two central tasks: QSP and the programming of operators.
QSP concerns the efficient loading of classical data or quantum states into a quantum register, which is a prerequisite for many quantum algorithms. 
QSP circuits can be constructed through optimization~\cite{Shirakawa2024}, analytically~\cite{An2023}, or deterministically using Tensor Networks (TNs)~\cite{Orus2019}.
In particular, Tensor Trains (TTs)~\cite{Oseledets2011} --one-dimensional TNs-- enable the efficient representation of weakly correlated data~\cite{Gourianov2022,Kiffner2023,VanHuelst2026,Michailidis2025}, which can then be translated into quantum gate sequences~\cite{Lubasch2020,Ran2020,Lin2021,Zhou2021,Rudolph2024,Szoldra2026}.
The programming of operators constitutes the other critical frontier in QCS, including, for example, analytical algorithms~\cite{An2023}, or variational quantum compilation, which involves efficiently executable low-depth Parameterized Quantum Circuits (PQCs)~\cite{Barratt2021,Tepaske2023,Mansuroglu2023,Keever2023,Keever2024,Robertson2025,Zhang2024}. 
Along this route, \citeauthor{Barratt2021}~\cite{Barratt2021} propose a hybrid partitioning scheme to distribute large quantum systems into subsystems, focusing on current hardware, where the recombination happens classically.
Similarly, \citeauthor{Tepaske2023}~\cite{Tepaske2023} advance the design of bricklayer circuits to minimize circuit depth in Hamiltonian simulations.
With a particular focus on time evolution and Trotterization strategies, References~\cite{Mansuroglu2023,Keever2023,Keever2024,Robertson2025,Zhang2024} also address simulation of Hamiltonians, whereby Reference~\cite{Zhang2024} employs Machine-Learning (ML) strategies to optimize circuit compilation tailored for two-dimensional chips.  

As a more general alternative to PQCs, latest research determines the quantum gates directly~\cite{Causer2024,Martin2024,Gibbs2025,Kotil2024,Le2025,Termanova2024,Nibbi2024,Wei2022,Wei2023}.
This transformation process entails the additional challenge of ensuring the unitarity of the resulting gates, since --unlike when using PQCs-- this is not inherently guaranteed.
Such a solution based on TT is demonstrated in References~\cite{Causer2024,Martin2024,Gibbs2025} and involves TT-sweeping strategies combined with polar decomposition, whereas \citeauthor{Gibbs2025}~\cite{Gibbs2025} incorporate the adaptivity for two-dimensional chips. 
Alternatively, other approaches~\cite{Kotil2024,Le2025,Termanova2024} leverage TT-based Riemannian optimization to build upon the underlying manifold's structure of unitary operators. 
Notably, the former works demonstrate a viable alternative to Trotterization for resource optimization by effective compression using TTs~\cite{Causer2024,Martin2024,Gibbs2025,Kotil2024,Le2025,Termanova2024}.
In addition, \citeauthor{Termanova2024}~\cite{Termanova2024} apply Riemannian optimization to engineering problems, thereby demonstrating the versatility and broad applicability of such approaches beyond quantum mechanics. 
In Reference~\cite{Termanova2024}, non-unitary dynamics are embedded into larger unitaries within the optimization procedure; this represents an alternative to the block-encoding technique for handling non-unitary operators, as presented in \citeauthor{Nibbi2024}~\cite{Nibbi2024}.

Despite the efficient realization of shallow circuits by one-dimensional TN models \cite{Termanova2024}, their limited ability to capture the long-range correlations of the operator restricts their applicability in scientific computing or ML \cite{Rudolph2024}.
To address these limitations, \citeauthor{Nibbi2024}~\cite{Nibbi2024} propose employing higher-dimensional tensor networks that better match the two-dimensional connectivity of certain quantum hardware architectures. 
In this regard, \citeauthor{Wei2022}~\cite{Wei2022,Wei2023} advance QCS techniques for two-dimensional quantum chips by employing higher-dimensional TNs, although the preparation time grows exponentially with system size \cite{Wei2022}. 
Integrating gate-sets to Riemannian optimization is introduced by \citeauthor{Guo2025}~\cite{Guo2025}, who utilize brick-wall circuit layouts in conjunction with TTs for real one-dimensional linear nearest neighbor architectures. 
Notably, variational QCS for quantum hardware with higher-dimensional connectivity or long-range programmable interactions -- often realized with neutral atoms~\cite{Evered2023} and trapped ions~\cite{Debnath2016} -- remains relatively unexplored, and the identification of scalable classical compression approaches is still an open problem.

%% OUR WORK
Within this work, we explore QCS in a way that allows certain characteristics of quantum hardware to be taken into account during the algorithmic design, specifically by restricting the qubit support of each  trainable gate to a user-specified connectivity list (cf. Section~\ref{sec:comp}).
To this end, we develop a backpropagation-based operator learning algorithm that accommodates a given qubit connectivity via register permutations. 
The algorithm itself is thus hardware-agnostic, whereas each learned circuit is specific to the supplied topology (cf.~Section~\ref{sec:opt}). 
The approach is demonstrated for one-dimensional and two-dimensional topologies in Section~\ref{sec:result} and can incorporate the chip layout as well as long-range and multi-body programmable interactions into the development of optimal quantum gates.
Therefore, the current implementation operates in the full operator space and does not rely on PQC or TT approaches. 
Non-unitary operators are block-encoded, similar to References~\cite{Termanova2024,Nibbi2024}, but only with one additional ancilla qubit.
Gate sequences are optimized by combining an automatic differentiation-based steepest descent approach with a projection technique based on singular-value decomposition, as it has been introduced by \citeauthor{Zhou2021}~\cite{Zhou2021}.
This approach is inspired by the tensor network-based optimization algorithms from Evenbly and Vidal~\cite{Evenbly2009} and presents an alternative to Riemannian optimization~\cite{Termanova2024}. 
To homogenize the accuracy across the resulting operators, we introduce an error-based regularization term into the learning procedure, which effectively reduces error outliers. 
The method can be extended to TN representations using standard techniques, e.g., Reference~\cite{Gibbs2025}, while preserving the optimization features of the full-space approach.

\smallskip
The article is organized as follows. 
The mathematical framework, the probabilistic features, and the computational strategy are described in Sections~\ref{sec:math} and \ref{sec:comp}.  
Section~\ref{sec:comp} also addresses the optimization strategy and analyzes the complexity of the entire framework. 
Section~\ref{sec:result} presents four applications, from quantum mechanics, quantum chemistry, and engineering sciences. 
A summary and conclusions are presented in Section~\ref{sec:conclusions}.
We denote arbitrary discrete vectors and matrices by Latin letters, and normalized vectors (quantum states) are indicated by the Dirac (bra-ket) notation. 
For representations of quantum systems, we employ the \textit{big-endian} convention.
The quantum applications are based on second quantization, whereas the engineering applications employ amplitude encoding.

%%%%%%%%%%%%%%%%%%%%%%%%%%%%%%%%%%%%%%%%%%%%%%%%%%%%%%%%%%%%%%%%%%%%
%%  METHODOLOGY
%%%%%%%%%%%%%%%%%%%%%%%%%%%%%%%%%%%%%%%%%%%%%%%%%%%%%%%%%%%%%%%%%%%%
\section{Methodology}
\label{sec:methodology}
\begin{figure}[t]
     \centering
     \includegraphics[width=0.95\textwidth]{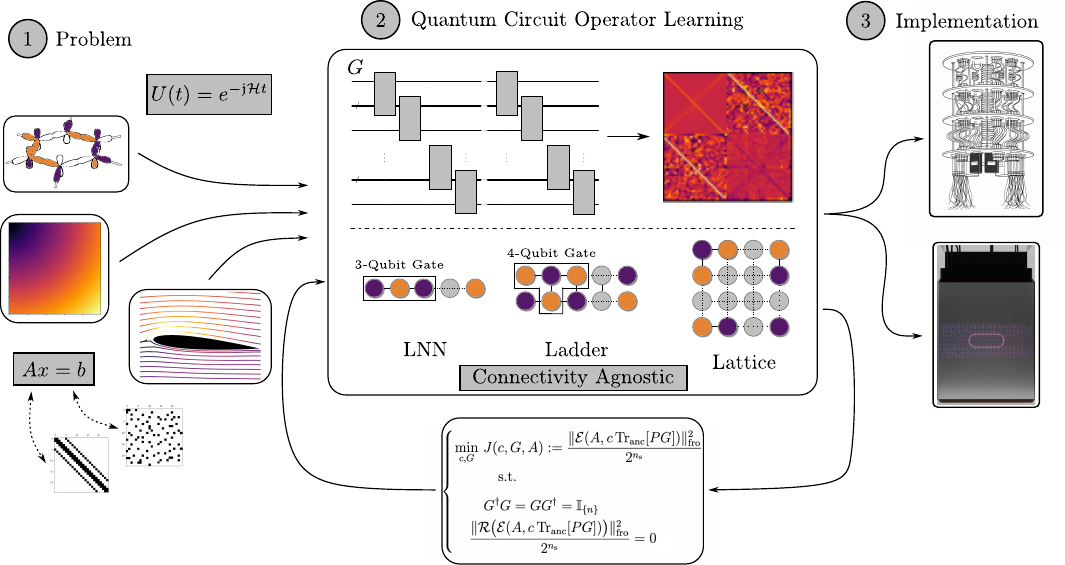}
     \caption{Illustration of the entire quantum operator learning approach.}
     \label{fig:overview}
 \end{figure}

In the following, we present the mathematical and computational framework that forms the basis of the proposed methodology for mapping discrete, potentially non-square operators onto quantum hardware. 
Learned operators obtained within this framework can be employed in a variety of quantum computing tasks, including quantum signal processing~\cite{Nibbi2024}, quantum linear solvers~\cite{Harrow2009} as an alternative to Trotterized time-evolution circuits~\cite{Suzuki1990}, incorporations into variational quantum algorithms~\cite{Jaksch2023,Over2024a,Bengoechea2025a}, and fault-tolerant algorithmic frameworks~\cite{Brearley2024,Over2024b,Bengoechea2025b}.
To this end, we consider a generic problem and outline a strategy to derive a gate-level representation optimized for the target quantum hardware through efficient operator learning.
An overview of the programming pipeline is illustrated in Figure~\ref{fig:overview}.

The synthesis begins with discretizing the target problem (cf. Figure~\ref{fig:overview}:~\textcircled{\raisebox{-.9pt} {1}}), arising in application fields such as quantum mechanics, quantum chemistry, or engineering. 
The discretization yields an operator ${A}$, e.g., representing a linear system ${A}{x}={b}$ or a quantum time-evolution operator $A \approx U(t) =  e^{-\mathfrak{j}\mathcal{H}t}$. 
Since ${A}$ is generally non-unitary, it is represented on quantum hardware via block-encoding into a larger unitary operator (cf. Section~\ref{sec:math}).
The encoding of ${A}$ into a quantum circuit is strongly influenced by hardware features (cf. Figure~\ref{fig:overview}:~\textcircled{\raisebox{-.9pt} {2}}), including chip architecture, qubit connectivity, gate size, and decoherence time. 
Current devices typically feature one- or two-dimensional qubit layouts with restricted connectivity, such as Linear Nearest-Neighbor (LNN), ladder, or square-lattice topologies.

\smallskip
The contribution presented here covers chip interconnection and layout, which are incorporated into the development of optimal quantum gates. 
These differences are illustrated by the example in Figure~\ref{fig:overview}, which depicts a 1D configuration (LNN) and a corresponding 3-qubit gate implementation of nearest neighbor interactions. 
Next to it, the ladder structure with a 4-qubit gate, relying also exclusively on nearest-neighbor interactions, exhibits comparable maximum information path lengths. 
In the case of an ideal, full inter-connectivity (all-to-all connectivity) \cite{Bluvstein2025}, arbitrarily complex, i.e., arbitrarily large, qubit gates can be represented through local 2-qubit nearest-neighbor interactions.
Notably, very few studies have addressed this aspect to date. 
Most studies are based on a simple 1D LNN configuration, which does not always reflect the reality of existing hardware. 
The method presented here aims to facilitate the development of optimal quantum gates for given hardware configurations. 
To this end, the study also investigates whether this significantly more flexible approach offers advantages in capturing correlations and structures. 

\smallskip
In the following, connectivity generally relates to the topology, the chip inter-connection, as well as short- or long-range qubit interactions. 
The cost function $J$ quantifies the approximation of the implemented operator $G$ to the target operator $A$, which is approximated by a sequence of fixed-body qubit gates assembled as a sequential ansatz. 
The range of the gates can thereby vary from nearest neighbor to long-range interaction.
In the final step (cf. Figure~\ref{fig:overview}:~\textcircled{\raisebox{-.9pt} {3}}), the derived gate sequence is brought to the target hardware, e.g., being superconducting or ion-trap machines.
Specific considerations of individual steps within the operator learning protocol follow in the next sections. 
A detailed description of the mathematical framework is given in Section~\ref{sec:math}.
The learning strategy, the sequential ansatz, and the optimization approach are part of the computational strategy, which is discussed and exploited along the algorithm in Section~\ref{sec:comp}. 

%%%%%%%%%%%%%%%%%%%%%%%%%%%%%%%%%%%%%%%%%%%%%%%%%%%%%%%%%%%%%%%%%%%%
%%  MATH
%%%%%%%%%%%%%%%%%%%%%%%%%%%%%%%%%%%%%%%%%%%%%%%%%%%%%%%%%%%%%%%%%%%%
\subsection{Mathematical Framework}
\label{sec:math}
We consider a problem described by a discrete operator ${A} \in \mathbb{C}^{N \times M} : N, M \in \mathbb{N},$ with an arbitrary structure. 
As a preprocessing step, the non-square operator is embedded into a larger square operator by setting $N = M = \max(N, M)$, where additional entries are filled with zeros. 
To implement the operator on a quantum device, at least $n_\text{s}=\lceil \log_2 N \rceil$ system qubits are required, so that the dimension of the qubit Hilbert space, $2^{n_\text{s}}$, is at least as large as the operator dimension $N$.
The number of qubits required depends on the encoding, and additional qubits are needed when the operator is not unitary ($A^\dagger A \neq  \mathbb{I} \neq A A^\dagger$). 
More precisely, a non-unitary target operator ${A}$ requires an embedding into a larger vector space, a procedure known as block encoding~\cite{Camps2020}. 
Block encoding is generally characterized by the normalization factor $c$, the number of ancillary qubits $n_\text{a}$, and an approximation error $\tau$. 
The block-encoding on the full register $n=n_\text{s}+n_\text{a}$ is defined as
\begin{equation}
\big \|
A - c \,
\text{Tr}_{\text{anc}}[P G] \,
\big \|_\mathrm{fro}
\leq \tau\,, \quad P = |0^{n_\text{a}}\rangle \langle 0^{n_\text{a}}| \otimes \mathbb{I}_{\{n_\text{s}\}}\, ,
\label{eq:block_encoding}
\end{equation}
where ${G}\in\mathbb{C}^{2^{n}\times2^{n}}$ is a unitary operator, $\mathbb{I}_{\{n_\text{s}\}}$ is the identity matrix of size $2^{n_\text{s}}\times 2^{n_\text{s}}$, $\text{Tr}_{\text{anc}}[\,\cdot\,]$ is the partial trace over the ancilla sub-system and $\|\cdot\|_\mathrm{fro}$ indicates the Frobenius norm.
The complete quantum register $|0^{n_\text{a}}\rangle \otimes \ket{\psi}$ consists of the ancilla system ($|0^{n_\text{a}}\rangle$) and the computational state ($\ket{\psi}$), and the target operator is encoded into the space prescribed by the projector $P$. 
Since $G$ is unitary, its block $\mathrm{Tr}_\mathrm{anc}[PG]$ must be a contraction, i.e., its largest singular value cannot exceed unity. 
This condition is ensured by the normalization factor $c \geq c_\mathrm{ideal}$, where $c_\mathrm{ideal} = \sigma_\mathrm{max}(A) = \|A\|_2$ denotes the spectral norm of $A$, such that $A/c_\mathrm{ideal}$ has a largest singular value of exactly one.
In the absence of approximation errors ($\tau = 0$), an exact encoding $\mathrm{Tr}_\mathrm{anc}[PG] = A/c$ is possible for any $c \geq c_\mathrm{ideal}$, with $c = c_\mathrm{ideal}$ being the smallest admissible choice.
The normalization factor, however, suppresses the post-selection success probability since the approximation $A \approx c\, \text{Tr}_\text{anc}[PG]$ implies $\|\text{Tr}_\text{anc}[PG]\,\ket{\psi}\|^2 \propto c^{-2}$~\cite{Gilyen2019}, where $\|\cdot\|$ denotes the Euclidean norm of a vector.
Therefore, a small normalization factor $c\approx c_\text{ideal}$ is desirable, which motivates the stabilizing term introduced in Equation~\eqref{eq:control_problem_unconstrained}.
Since the success probability also depends on the state \cite{Gilyen2019}, we report its average over all input states, i.e.,
\begin{equation}
        p(PGP) =  \frac{\|PGP\|_\mathrm{fro}^2}{2^{n_\text{s}}}.
        \label{eq:succ_prob_avg}
\end{equation}
If ${A}$ is unitary, as in the case of a Hamiltonian simulation, the block encoding using ancillas is not required ($n_\text{a}=0$), the projector ${P}$ is the identity matrix $\mathbb{I}_{\{n\}}$,  $c=c_\text{ideal}=1$, and $p(A)=p(G) = 1$.
\bigskip

In this work, we consider a single-qubit block encoding $n_\text{a} = 1$. 
We note that our procedure can be realized with any number of qubits, unlike other approaches~\cite{Termanova2024, Nibbi2024} which require a problem-dependent number of ancilla qubits. 
We assume that ${G}$ is a sequence of parameterized quantum gates expressed by a generic ansatz introduced in Section~\ref{sec:comp}.
We aim to find solutions to Equation~\eqref{eq:block_encoding} with small truncation error $\tau$.
For this reason, we define the residual for two matrices $A$, $B$ as
\begin{equation}
    {\mathcal{E}}(A,B) = A - B
    \label{eq:error_rel}
\end{equation}
and the relative error as
\begin{equation}
\label{eq:relativeErrorScalar}
        \varepsilon(A,B)= \frac{\|{\mathcal{E}}(A,B)\|_\mathrm{fro} }{\|A\|_\mathrm{fro}}\, .
\end{equation}
Our approach optimizes only the average error. 
Norms that capture the worst-case error, such as the spectral norm, generally require iterative procedures within tensor-network implementations, whose efficiency and convergence cannot be guaranteed at the scale required for large systems.
Nevertheless, Appendix~\ref{app:specfro} discusses the worst-case error, as measured by the spectral norm, for the numerical results presented in this study.

\bigskip
The optimization problem for learning the target operator ${A}$ is defined as
\begin{subnumcases}{}
    \label{eq:control_problem}
    \min_{c,{G}}\, J(c,{G},A):= \frac{\|\mathcal{E}(A,c \,
\text{Tr}_{\text{anc}}[P G])\|_{\mathrm{fro}}^2}{2^{n_\text{s}}}\, \\
    \qquad \quad \text{s.t.} \notag \\
    \label{eq:control_problem_unitary}
    \qquad{G}^\dagger {G}={G}{G}^\dagger=\mathbb{I}_{\{n\}} \, \\ 
    \label{eq:control_problem_regularization}
    \quad  \frac{\|{\mathcal{R}}\big(\mathcal{E}(A,c \,
\text{Tr}_{\text{anc}}[P G])\big)\|_\mathrm{fro}^2}{2^{n_\text{s}}} = 0\, ,
\end{subnumcases} 
where ${2^{n_\text{s}}}$ realizes a normalization, making sure that the objective remains finite with increasing system sizes, and ${\mathcal{R}}$ is a variant of $\ell_2$-regularization.

The purpose of regularization is to limit error outliers, maintaining constant error magnitudes throughout the optimization, or in other terms, that a non-zero residual $\mathcal{E}$ should be sufficiently smooth.
This is achieved by penalizing the objective with a filter contribution evaluated for the entries of the matrix ${\mathcal{E}}$. 
The filter is defined as a five-point stencil on two dimensions with periodic boundary conditions, such that the regularization term defined at location $ij$ is
\begin{equation}
    {\mathcal{R}}({\mathcal{E}})_{ij} = (\mathcal{E}_{i-1j}+\mathcal{E}_{ij-1} -4 \mathcal{E}_{ij} + \mathcal{E}_{i+1j} + \mathcal{E}_{ij+1})_{ij} \, .
    \label{eq:regu}
\end{equation}

For implementation, the constrained problem~\hyperref[eq:control_problem]{{(5)}} is reformulated as an unconstrained one. 
First, the evolution of a closed quantum system is unitary. 
Each gate in the circuit must therefore correspond to a unitary operator, which in turn ensures that the full circuit is unitary, as expressed in Equation~\eqref{eq:control_problem_unitary}.
We enforce this unitary constraint on the individual gates by combining a relaxation strategy with a retraction onto the unitary manifold, as described in detail in Section~\ref{sec:comp}.
Second, the regularization~\eqref{eq:control_problem_regularization} is handled by a penalty approach, incorporating the constant penalty term $\rho\in\mathbb{R}^+$ \cite{Nocedal2006}. 
As a result, the unconstrained optimization problem achieves optimality with respect to its contributions, and is explicitly described by
\begin{equation}
     \min_{c,{G}}\,\mathcal{J}(c,{G},A,\rho,\mu) := \frac{\|\mathcal{E}(A,c \,
\text{Tr}_{\text{anc}}[P G])\|^2_{\mathrm{fro}}}{2^{n_\text{s}}} + \rho  \frac{\|{\mathcal{R}}\big(\mathcal{E}(A,c \,
\text{Tr}_{\text{anc}}[P G])\big)\|^2_\mathrm{fro}}{2^{n_\text{s}}} \,+ \mu {c^2} \,.
     \label{eq:control_problem_unconstrained}
\end{equation}
The additional term $\mu c^2$, with weight $\mu \in \mathbb{R}^+$, penalizes large normalization factors.
Without this term, the optimization can converge to solutions with $c \gg c_\mathrm{ideal}$.
An example is presented in Appendix~\ref{app:influence_c}.
For a suitably chosen $\mu$, we did not observe a significant loss in approximation accuracy, while such solutions are avoided.

The solution of Equation~\eqref{eq:control_problem_unconstrained} is characterized by a stationary point, given by optimal controls $c^*,\,{G}^*$.
At this point, the partial derivatives of $\mathcal{J}$ w.r.t the control vanish for all variations $\delta c$, and the admissible variations $\delta {G}$ at $G$, viz.,
\begin{equation}
    \label{eq:saddle}
    \frac{\partial \mathcal{J}}{\partial c} \delta c=0\, \; \forall \delta c \, ,\quad 
    (\nabla_\text{R} \mathcal{J}, \delta G)=0 \, \; \forall  \delta G \in T_G\,U(2^n)\,,
\end{equation}
with $(\cdot,\cdot)$ being the respective inner product of the underlying unitary manifold $U(2^n)$, and $\nabla_\text{R}$ the Riemann gradient, and $T_G\,U(2^n)$ is the
tangent space at $G$ \cite{Absil2008,Luchnikov2021}. 
We employ a retraction based on the singular value decomposition (SVD) (cf. Algorithm~\ref{alg:opt}), enforcing the constraint in Equation~\eqref{eq:control_problem_unitary}.
This system of first-order optimality conditions can be directly associated with each contribution to the cost function.
For example, the solution for $c$ is obtained by imposing the first-order optimality condition associated with the quadratic cost function defined in Equation~\eqref{eq:control_problem_unconstrained} and solving Equation~\eqref{eq:saddle} w.r.t $c$.
This results in an analytical expression for the normalization constant, which is
\begin{equation}
    c^* = \text{Re}\Bigg(
      \frac{\text{Tr}_\text{sys}\big[A^\dagger
            \text{Tr}_{\text{anc}}[P G]\big]
          + \rho\,\text{Tr}_\text{sys}\big[
            \mathcal{R}(A)^\dagger\,
            \mathcal{R}\big(\text{Tr}_{\text{anc}}[P G]\big)\big]}{
            2^{n_\text{s}}\mu + \|P G P\|^2_\mathrm{fro}
          + \rho\,\big\|\mathcal{R}\big(
            \text{Tr}_{\text{anc}}[P G]\big)\big\|^2_\mathrm{fro}}
      \Bigg)\,,
      \label{eq:c_sol}
\end{equation}
where $\text{Re}(\cdot)$ indicates the real-part. 
As evident from the equation above, the optimal $c^*$ depends directly on $G$ and can therefore be eliminated as a degree of freedom.
The computational treatment of first-order optimality is further addressed in Section~\ref{sec:opt}.
An investigation of the performance of the regularizer ${\mathcal{R}}$ is part of the numerical studies, presented in Section~\ref{sec:result}.

%%%%%%%%%%%%%%%%%%%%%%%%%%%%%%%%%%%%%%%%%%%%%%%%%%%%%%%%%%%%%%%%%%%%
%%  COMP. STRAT
%%%%%%%%%%%%%%%%%%%%%%%%%%%%%%%%%%%%%%%%%%%%%%%%%%%%%%%%%%%%%%%%%%%%
\subsection{Computational Strategy} 
\label{sec:comp}
Building on the mathematical framework introduced above, this section presents the computational implementation of the proposed operator learning approach. 
We employ a hybrid strategy in which gate primitives are trained on classical hardware and the resulting operators are prepared for subsequent realization on quantum devices.
Accordingly, the focus here is on the treatment of the classical optimization problem, with particular emphasis on the hierarchical optimization scheduler of Section~\ref{sec:opt} and the algorithm's complexity specifications of Section~\ref{sec:complexity}.

\begin{figure}[t]
    \centering
    \includegraphics[width=0.85\textwidth]{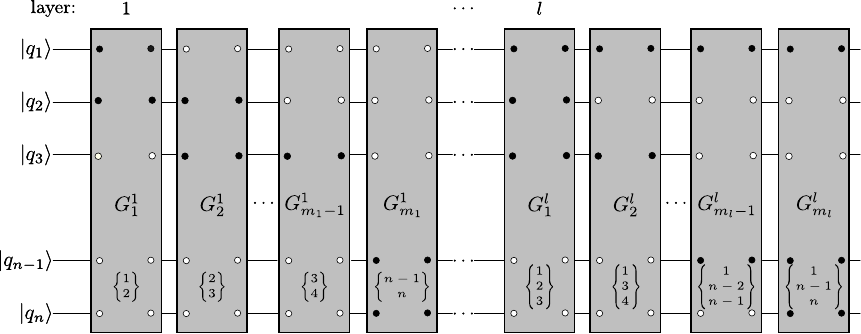}
    \caption{Illustration of an $l$-layer trainable quantum circuit ansatz $G$, cf. Equation~\eqref{eq:assembleG}.
    For each gate, the subset of qubits on which it acts non-trivially, denoted by $\mathcal{G}_i^j$, is indicated by the closed vertices, or equivalently by its connectivity list illustrated with curly brackets.
    The first layer is composed of $m^1=n-1$ nearest-neighbor two-qubit gates, whereas the $l$-th layer consists of $m^l=n-2$ three-qubit gates, of which all but the first are long-range gates.}
    \label{fig:QuantumAnsatz}
\end{figure}

\smallskip
Since our algorithm is hardware-agnostic, we employ as a default ansatz a general circuit without constraints in the layout, illustrated in Figure~\ref{fig:QuantumAnsatz}. 
Thereby, we ensure to fit any hardware into the same computing layout, which allows adapting to connectivity, long-range quantum gates, or other device-specific characteristics. 
Furthermore, we consider a standard $n$-qubit register $\ket{q}=\ket{q_1}\otimes \cdots \otimes \ket{q_n}\in \mathbb{C}^{2^n}\, $ and the trainable quantum circuit defined as
\begin{equation}
G=\prod_{j=1}^{l}\prod_{i=1}^{m^j} G_i^j \,,
    \label{eq:assembleG}
\end{equation}
where the gates are organized into $l$ layers, each containing $m^j$ gates.
Each trainable gate $G_i^j$ is a unitary operator, satisfying the unitary condition $(G_i^j)^\dagger G_i^j=\mathbb{I}\,,$ where $\mathbb{I}$ is the respective identity matrix, and that acts non-trivially only on a subset of qubits $\mathcal{G}_i^j=\{g_1,\ldots,g_r\}\subseteq \{1,\ldots,n\}\,$. 

More precisely, $G_i^j$ is the canonical embedding into the full $n$-qubit Hilbert space of a local operator $\widetilde{G}_i^j$ acting on the $r^j$ qubits indexed by $\mathcal{G}_i^j$, while acting as the identity on all remaining qubits.
The gate $G_i^j$ is often referred to as an $r^j$-body gate. 
A further classification of the gates is provided by their range, defined as the maximum distance between any two qubits in the support $\mathcal{G}_i^j$.
This distance is measured with respect to the placement of the qubits in the underlying hardware architecture.
Consequently, the gate sizes may range from two-qubit gates up to many-body gates acting on the entire register.
To this end, the variables $l$, $m^j$ and $r^j$ are explicitly set with $\mathcal{G}_i^j$ prescribed by the user.
Thereby, different connectivity patterns can be straightforwardly implemented to model, e.g., long-range interactions or two-dimensional lattice structures. 
For example, the gate $G_1^1 = \widetilde{G}^1_1 \otimes \mathbb{I}_3 \otimes \ldots \otimes\mathbb{I}_n$ is illustrated in Figure~\ref{fig:QuantumAnsatz} where $\mathbb{I}_i\in \mathbb{R}^{2 \times 2}$ is the identity matrix associated to qubit $i$.
The gate $G_1^1$ acts non-trivially only on the first two qubits of the register, corresponding to the subset $\mathcal{G}_1^1=\{1,2\}$. 
Therefore, it is a two-body gate ($r^1=2$), and for a linear hardware layout following the qubit enumeration, it is a range-2 (or nearest-neighbor) gate.
Instead, the last gate $G_{m^l}^l$ acts on the first and last two qubits of the register given by the subset   $\mathcal{G}_{m^l}^l=\{1,n-1,n\}$, which makes it an $r^l$-qubit gate ($r^l=3$), and for linear hardware, therefore a long-range interacting gate.
The number of gates $m^j$ in each layer and the number of qubit supports $r^j$, i.e., the number of qubits each gate acts on (assumed to be uniform within a layer) can be chosen independently for each layer.
This allows the approach to be adapted to external constraints, such as the connectivity or native gate sizes of a given hardware platform.

In the following section, we describe how the trainable quantum circuit $G$ is learned by optimizing the non-trivial component of each trainable gate.

%%%%%%%%%%%%%%%%%%%%%%%%%%%%%%%%%%%%%%%%%%%%%%%%%%%%%%%%%%%%%%%%%%%%
%%  OPT
%%%%%%%%%%%%%%%%%%%%%%%%%%%%%%%%%%%%%%%%%%%%%%%%%%%%%%%%%%%%%%%%%%%%
\subsubsection{Strategy}
\label{sec:opt}
This section focuses on the optimization strategy by which the gates $G$ -- assembled as introduced in the previous section -- are iteratively updated to minimize the objective $\mathcal{J}$. 
Rather than utilizing Riemannian gradient~\cite{Luchnikov2021}, we adopt automatic differentiation~\cite{Zhou2021} that can be naturally extended to TN representations~\cite{Liao2019}. 
To this end, the presented approach employs a gradient-based optimizer, where the derivatives are approximated by backpropagation through the computational graph -- involving the computation of the environment tensor -- as it is common in ML~\cite{Rumelhart1986}. 

To reduce the computational cost, we perform the optimization at the level of the local gate primitives $\widetilde{G}_i^j$, rather than using their full Hilbert-space embeddings $G_i^j$.
This avoids manipulating exponentially large matrices whenever only a few qubits are involved, leading to a substantial reduction in computational overhead (cf. Section~\ref{sec:complexity}).
For clarity of presentation, however, we formulate the algorithm below in terms of the embedded gates $G_i^j$. 
We refer the reader to Reference~\cite{Gibbs2025} for the efficient gradient-evaluation procedure based on the non-embedded gates. 
This modification affects only the implementation of the gradient computation, and the structure of the remaining algorithm stays unchanged.

The update of the local gate primitives then follows gate-wise in the steepest descent direction as 
\begin{equation}
    \bar{G}_i^j =  \widetilde{G}_i^j - \eta  \frac{\partial  \mathcal{J} }{\partial \widetilde{G}_i^j} \qquad \forall \, i  \in [1,m^j],\,\, j\ \in[1,l] \, ,
\end{equation}
where $\bar{G}_i^j$ are non-unitary local latent gates, $\eta$ is the learning rate (stepsize), and ${\partial  \mathcal{J} }/{\partial  \widetilde{G}_i^j}$ is the Euclidean derivative. 
Using the Euclidean update violates the unitary constraint in Equation~\eqref{eq:control_problem_unitary}, such that all $\bar{G}_i^j$ require to be corrected.
This is done by a retraction, realized by performing an SVD on each latent gate primitive, viz.,
\begin{equation}
     \text{SVD}\big(\bar{G}_i^j \big) = U \Sigma V^\dagger \quad \rightarrow \quad 
   \widetilde{G}_i^j = U V^\dagger  \,,
\end{equation}
where we replace the diagonal matrix of singular values $\Sigma$ with the identity $\mathbb{I}\in \mathbb{R}^{2^{r^j} \times 2^{r^j}}$, retracting the latent gate back onto the unitary group $U(2^{r^j})$.
This retraction is an approximation to the Riemannian exponential map~\cite{Absil2008}.
Although this retraction of the operator back onto the manifold of the unitary group introduces an error,
$\tilde{G}_i^j$ is, by construction, the unitary matrix closest to $\bar{G}_i^j$ in the Frobenius norm~\cite{FanHoffman1955}.
More specifically, for a gate acting on $r^j$ qubits, $\widetilde{G}_i^j \in U(2^{r^j})$. Since \(\min_{G\in U(2^{r^j})}\|e^{\mathfrak{j}\phi}\bar G- G\|_\mathrm{fro}^2=\min_{G\in U(2^{r^j})}\|\bar G- G\|_\mathrm{fro}^2\),  the gate’s global phase $\phi\in\mathbb{R}$ does not affect the retraction error.
The approach can be understood as a polar decomposition and is inspired by the Evenbly-Vidal algorithm \cite{Evenbly2009}. 
For quantum state preparation, this approach has been recently applied in Reference~\cite{Zhou2021}, where \citeauthor{Szoldra2026}~\cite{Szoldra2026} highlighted improvements over Riemannian descent methods, further motivating the protocol presented below.

With the corrected local gates, the operator programming protocol first canonically embeds the local gate primitives $\widetilde{G}_i^j$ ($r^j$-qubit gate) into the corresponding $n$-qubit Hilbert space, resulting in $\widehat{G}_i^j$. 
Secondly, a quantum permutation $S(\mathcal{G}_i^j)$ is applied to align with the given connectivity and hardware support $\mathcal{G}_i^j$, viz., $S(\mathcal{G}_i^j)^\dagger \widehat{G}_i^j S(\mathcal{G}_i^j)={G}_i^j$.
In a last step, the gate $G$ results from Equation~\eqref{eq:assembleG}, which allows to compute the normalization $c$ and to finally evaluate the cost-function value $\mathcal{J}$. 

The optimization procedure is subordinate to a hierarchical optimization approach, which is based on prioritizing the training of the most significant qubits first. 
A differential relevance among individual qubits is particularly present using amplitude encoding, commonly employed for running scientific computing problems on QCs.
The hierarchical optimization procedure is detailed in Algorithm~\ref{alg:strat} and builds upon Reference~\cite{Gharibyan2024}.
\begin{algorithm}[H]
\caption{Hierarchical optimization protocol.}
\label{alg:strat}
\begin{algorithmic}[1]
\REQUIRE $l$, \texttt{tol}, \texttt{maxit}, $\mathcal{G}_i^j$
\STATE $m^j \leftarrow \mathcal{G}_i^j$
  \COMMENT{Initialize gate counts}
\FOR{$j=1$ to $l$}
    \FOR{$i=1$ to $\min_{k=1,\ldots,j} m^k$}
        \STATE Optimize $(\widetilde G_i^1,\ldots,\widetilde G_i^j)$
        by executing Algorithm~\ref{alg:opt}
        \label{alg:strat:op1}
    \ENDFOR
    \STATE Jointly optimize
    $\{\widetilde G_i^k:\,1\leq k\leq j,\;1\leq i\leq m^k\}$
    by executing Algorithm~\ref{alg:opt}
    \label{alg:strat:op2}
\ENDFOR
\STATE Train all gates
$\{\widetilde G_i^j:\,1\leq j\leq l,\;1\leq i\leq m^j\}$
by executing Algorithm~\ref{alg:opt}
\label{alg:strat:op3}
\end{algorithmic}
\end{algorithm}
Algorithm~\ref{alg:strat} builds up the circuit layer by layer.
For each layer $j$, the gates with index $i$ are first optimized together with the gates of the same index in all preceding layers (cf. Algorithm~\ref{alg:strat} -- line~\ref{alg:strat:op1}),
before all gates of layers $1$ to $j$ are optimized jointly (cf. Algorithm~\ref{alg:strat} -- line~\ref{alg:strat:op2}).
Once the last layer has been added, all gates are trained in a final global optimization
(cf. Algorithm~\ref{alg:strat} -- line~\ref{alg:strat:op3}).
Without the hierarchical stages, only the final global optimization is performed, cf. Algorithm~\ref{alg:strat} -- line~\ref{alg:strat:op3}.
Throughout, gates that have already been optimized but are not part of the current step are held fixed.
To reduce computational effort, all gates that have not yet been optimized are excluded from the evaluation, so that the operator learning protocol starts with a one-gate quantum circuit.
Within each call of Algorithm~\ref{alg:opt} (cf. Algorithm~\ref{alg:strat} -- line~\ref{alg:strat:op1}, \ref{alg:strat:op2} and line~\ref{alg:strat:op3}), the optimization is started for a user-defined maximum number of iterations $\texttt{maxit}$ and is terminated based on a user-defined tolerance $\texttt{tol}$ on the objective's derivative. 
The underlying optimization procedure is summarized in Algorithm~\ref{alg:opt}.
 
For the gradient-based optimization denoted in Algorithm~\ref{alg:opt} -- line~\ref{alg:opt:grad}, we employ two algorithms, which are either the L-BFGS~\cite{Nocedal2006} or the ADAM~\cite{Kingma2015} algorithm.
The ADAM optimizer is particularly advantageous for enhancing Barren Plateau (BP) robustness, while problems that are more likely to be convex are addressed using the Quasi-Newton approach (L-BFGS), to reduce computational costs. 
In both cases, we incorporate informed initial guesses whenever available, otherwise, the method initializes the gates with identities. 
By omitting unused layers during backpropagation, we significantly reduce computational costs.
In combination with hierarchical learning, we observed that L-BFGS achieves higher optimization accuracy and restricted our implementation to this optimizer.
The following section (cf. Section~\ref{sec:complexity}) discusses in detail the computational complexity and scalability of the proposed protocol.

\begin{algorithm}[H]
   \caption{Optimization procedure used for efficient operator learning.}
   \label{alg:opt}
    \begin{algorithmic}[1]
       \REQUIRE $\rho$, $\mu$, $i_{\max}$, $j_{\max}$, ${A}$, $n_\text{s}$, \texttt{tol}, \texttt{maxit}, $\mathcal{G}_i^j$
        \IF {${A}^\dagger {A} = {A}{A}^\dagger =\mathbb{I}_{n_\text{s}}$}
            \STATE  $n_\text{a}=0$ and  ${P} =\mathbb{I}_{n_\text{s}}$ 
        \ELSE
            \STATE $n_\text{a}=1$ and  ${P}=|0^{n_\text{a}}\rangle \langle 0^{n_\text{a}}| \otimes \mathbb{I}_{n_\text{s}}$
       \ENDIF
       \STATE $k=0$
        \STATE Assemble ${S}(\mathcal{G}_i^j)$ based on connectivity list $\mathcal{G}_i^j$
   \WHILE{$\|{\partial  \mathcal{J} }/{\partial {G}}\|>\texttt{tol} \text{ and } k < \texttt{maxit}$}
        \STATE Canonically embed $\widetilde G_i^j$ as $\widehat{G}_i^j$
       \STATE Compute ${G_i^j}\!=\!{S}^\dagger(\mathcal{G}_i^j) \widehat{G}_i^j{S}(\mathcal{G}_i^j)$ \COMMENT{Apply connectivities}
      \STATE Compute $B\!=\!\text{Tr}_{\text{anc}}[P G]$
      \STATE Compute $c\!=\!\text{Re}\Big(\big(\text{Tr}_\text{sys}[A^\dagger B]
      + \rho\,\text{Tr}_\text{sys}[\mathcal{R}(A)^\dagger \mathcal{R}(B)]\big)
      / \big(\mu 2^{n_\text{s}} + \|B\|^2_\mathrm{fro}
      + \rho\,\|\mathcal{R}(B)\|^2_\mathrm{fro}\big)\Big)$
        \STATE Evaluate $\mathcal{J}\!=\!{\|\mathcal{E}(A,c \,B)\|^2_{\mathrm{fro}}}/{2^{n_\text{s}}} + \rho  {\|{\mathcal{R}}\big(\mathcal{E}(A,c \,B)\big)\|^2_\mathrm{fro}}/{2^{n_\text{s}}} \,+ \mu {c^2}$
        \FORALL{$G_i^j$, where $i\leq i_{\max}$ and $j\leq j_{\max}$}
            \STATE Compute required derivatives by backpropagation ${\partial  \mathcal{J} }/{\partial \widetilde G_i^j}$
            \STATE Apply Euclidean update ${\bar{G}}_i^j\!=\!\widetilde G_i^j - \eta\, {\partial  \mathcal{J} }/{\partial \widetilde G_i^j}$ using a local gradient based approach \label{alg:opt:grad}
            \STATE Enforce constraint \eqref{eq:control_problem_unitary}: $\text{SVD}({\bar{G}}_i^j)\!=\! U \Sigma V^\dagger \; \Rightarrow \;   \widetilde G_i^j = U V^\dagger $ \COMMENT{Retract onto $U(2^{r^j})$ manifold}
            \label{alg:opt:unitary}
        \ENDFOR
        \STATE $k = k +1$
    \ENDWHILE
    \end{algorithmic}
 \end{algorithm}

To improve BP resilience, we incorporate multiple strategies, comprehensively reviewed in Reference~\cite{Larocca2025}.
First, the hierarchical optimization algorithm presented in Algorithm~\ref{alg:strat} reduces the optimization's sensitivity by avoiding sub-optimal local minima encountered in non-convex optimization landscapes. 
This advantage is elaborated upon in Reference~\cite{Rudolph2024}, which implements a sophisticated strategy to mitigate BPs and achieves improved accuracy.
Second, the inclusion of developed regularization techniques shows promise in enhancing robustness against BP phenomena. 
Further options for resolving BP issues include optimized initial guesses, leveraging classical shadows~\cite{Sack2022}, or using classically motivated surrogate models~\cite{McClean2018,Friedrich22} closely related to this work. 

%%%%%%%%%%%%%%%%%%%%%%%%%%%%%%%%%%%%%%%%%%%%%%%%%%%%%%%%%%%%%%%%%%%%
%%  COMPLEXITIES
%%%%%%%%%%%%%%%%%%%%%%%%%%%%%%%%%%%%%%%%%%%%%%%%%%%%%%%%%%%%%%%%%%%%
\subsubsection{Complexity}
\label{sec:complexity}
This section analyzes the computational complexity of the proposed method. 
In particular, we focus on the dominant costs arising in Algorithms~\ref{alg:strat} and \ref{alg:opt}, and their implications for the subsequent quantum hardware implementation. 
For comparison with the existing literature, we consider the worst case of a constant number of gates $m = \max_j m^j$ per layer and a constant gate size $r = \max_j r^j$ throughout the circuit, which yields an upper bound on the resources required for the general ansatz.
Throughout, $r$ is assumed to be as small as the desired expressivity permits, in particular sufficiently smaller than $n$, since the cost of implementing an $r$-qubit gate grows exponentially with $r$.
Further, we assume the global optimization (cf. Algorithm~\ref{alg:strat} -- line~\ref{alg:strat:op3}) as a worst-case scenario since this algorithmic step incurs the highest computational costs. 
Therefore, we concentrate on estimating the cost of Algorithm~\ref{alg:opt}.

In the operator encoding protocol, each local gate primitive belonging to the unitary group $U{(2^r)}$ is parametrized by a $2^{r}\times2^{r}$ matrix with complex-valued entries.
The corresponding control vector $\lambda$ is constructed by concatenating the parameters of all gates, ordered layer-wise from top to bottom and, within each layer, from left to right, terminating at $G_l^{m}$. 
Since each matrix entry can be in general complex-valued, each entry is represented by two real parameters, i.e., $\lambda_i+\lambda_{i+1} \mathfrak{j}$, with $\mathfrak{j}$ being the imaginary unit.
Consequently, the number of parameters per gate results in $2^{2r+1}$. 
This number reduces by half when the target operator is restricted to real values. 
If we now assume $l$ layers with $m$ gates per layer and a complex-valued operator to be learned, the total number of degrees of freedom scales with $\mathcal{O}(l \, m\, 2^{2r+1})$. 

For evaluation of the cost function, these local gate primitives $\widetilde{G}_i^j$ are canonically embedded into the group $U{(2^{n})}$ to perform the matrix multiplication given in Equation~\eqref{eq:assembleG}.
This multiplication is subject to the computational cost of $\mathcal{O}(2^{3{n}})$. 
If local primitives $\widetilde{G}_i^j$ are used instead of embedded gates ${G}_i^j$, the contraction cost can be improved to $\mathcal{O}(2^{2{n}}\, 2^{r})$. 
As this contraction is performed for $l$ layers and $m$ gates the contraction complexity is $\mathcal{C}_\text{cont.} = \mathcal{O}(2^{2n} \, 2^{r} \,l \,m)$.

Another central property of the optimization procedure described in Algorithm~\ref{alg:opt} -- line~\ref{alg:opt:unitary} is the enforcement of the constraint~\eqref{eq:control_problem_unitary} by the retraction on the unitary manifold after each gradient-based update of the local gates $\widetilde{G}_i^j$. 
This retraction requires computing an SVD for each gate modification, with the computational effort for $m$ gates and $l$ layers scaling as $\mathcal{C}_\text{SVD} = \mathcal{O}( 2^{3r}\, l\,m )$.
Accounting for all contributions during the optimization, the total cost per $k$ iterations is 
\begin{equation}
        \mathcal{C}_\text{opt} = k \,  (\mathcal{C}_\text{SVD} + \mathcal{C}_\text{cont.}) =   k \, l \, m \, \mathcal{O}( 2^{3r} + 2^{2n} \, 2^{r} )\, .
    \label{eq:comp_opt}
\end{equation}
This computational effort is accumulated across the hierarchical optimization strategy (cf. Algorithm~\ref{alg:strat}). 
However, this increase is partially mitigated by reducing the degrees of freedom within early hierarchical steps, as discussed in Section~\ref{sec:opt}.

\smallskip
Equation~\eqref{eq:comp_opt} indicates an exponential scaling with the system size~$n$ and the gate size~$r$. 
The scaling with respect to $n$ is intrinsic to any classical approach that does not rely on TN. 
In TN implementations, the algorithm does not scale exponentially with the system size but instead polynomially with the complexity of the tensor network described by the auxiliary dimension~\cite{Gibbs2025}.
However, the presented protocol can rely on small multi-qubit gates and therefore can meet the hardware requirements.  Therefore, individual problems come with individual gate complexities, for which the computational resources are estimated by Equation~\eqref{eq:comp_opt}.

For estimating the gate complexity $\mathcal{C}_{gate}$, we assume that decomposing a general $r$-qubit gate into CNOTs requires on the order of $4^{r-1}$ gates~\cite{Shende2004}. 
The total CNOT count of the circuit is then approximately
\begin{equation}
     \mathcal{C}_\mathrm{gate} \approx \sum_{i,j} 4^{r-1}\, \text{gates}, 
    \label{eq:CNOT}
\end{equation}
which is shown in Figure~\ref{fig:ansatzcomplexity} for different values of $l$ and $r$. 
For a fixed or at most polylogarithmically growing number of layers $l$, this count scales polylogarithmically with the operator dimension $N$. 
Note that in this figure, the combination of $l=1\, \text{(solid)}$ with $r = 2 \, \text{(cross)}$ is equivalent in gate resources to a Suzuki--Trotter (ST) first-order decomposition, cf. Section~\ref{sec:Quantum}.
\begin{figure}[t]
    \centering
    \tikzsetnextfilename{AnsatzQubitDepth}
\begin{tikzpicture}
    \begin{semilogyaxis}[
    xlabel = {$n$},
    ylabel = {\# CNOT gates},
    ymajorgrids,
    yminorgrids,
    ymin=1,
    width=12cm,
    height=4.5cm,
    ymax=10^4,
    xmax=15,
    xmin=2,
    %height=0.3\linewidth,
    ]

    % (n-n_g+1)
    %n_g= 2  d=1
    \addplot[mark=x, domain=2:15, samples=14, mark options={scale=0.9,solid} ]{4*(x-2+1)};
    % \node[right] at (15,55) {\scriptsize{$l=1$}};

    %n_g= 2  d=3
    \addplot[mark=x, dashed, domain=2:15,samples=14, mark options={scale=0.9,solid}]{4*3*(x-2+1)};
    % \node[right] at (15,150) {\scriptsize{$l=3$}};

    %n_g= 2  d=5
    \addplot[mark=x, dotted, domain=2:15, samples=14,mark options={scale=0.9,solid}]{4*5*(x-2+1)};

    % \node[right] at (15,300) {\scriptsize{$l=5$}};

    %n_g= 3  d=1
    \addplot[mark=o,mark options={scale=0.9,solid}, domain=3:15, samples=13, mark options={scale=0.8,solid} ]{16*(x-3+1)};
     %n_g= 3  d=3
    \addplot[mark=o,mark options={scale=0.9,solid}, samples=13,dashed,domain=3:15, ]{16*3*(x-3+1)};
    % %n_g= 3  d=5
    \addplot[mark=o,mark options={scale=0.9,solid}, samples=13,dotted,domain=3:15, ]{16*5*(x-3+1)};
    
    %n_g= 4  d=1
    \addplot[mark=star,   samples=12,  domain=4:15,mark options={scale=0.9,solid} ]{64*(x-4+1)};
    %n_g= 4  d=3
    \addplot[mark=star,  dashed, samples=12 ,domain=4:15, mark options={scale=0.9,solid}]{64*3*(x-4+1)};
    %n_g= 4  d=5
    \addplot[mark=star,  dotted, samples=12,domain=4:15, mark options={scale=0.9,solid}]{64*5*(x-4+1)};

\addplot [thin,gray,domain=0:15,] {(2^x)};
\node[right,align=left] at (10.5,6000) {\textcolor{gray}{$\mathcal{O}(2^n)$}};

% \addplot [thin,gray,domain=1:15,]   {(x^2)};
% \node[right,align=left] at (7.9,220) {\textcolor{gray}{$\mathcal{O}(n^2)$}};

\end{semilogyaxis}
\end{tikzpicture}
    \caption{Gate complexity $\mathcal{C}_{gate}$ over number of qubits for layers $l=\{1$ (solid), {$3$ (dashed)}, $5 \text{ (dotted)}\}$ (line-type) and gate sizes $r=\{2\, (\times),\, 3\, (\circ),\,4\, (\star)\}$ (symbol), cf. Equation~\eqref{eq:CNOT}.}
    \label{fig:ansatzcomplexity}
\end{figure}

\smallskip
Considering non-unitary operations, the approaches in References~\cite{Termanova2024,Nibbi2024} require multiple ancilla qubits, depending on the auxiliary dimension of the employed tensor network approach.
Alternatively, the Linear Combination of Unitaries (LCU) algorithm unitary decomposes a matrix $A$ in $\kappa$ unitaries employing $n_\text{a} = \lceil \log_2(\kappa) \rceil$ ancillas. 
Instead, the presented framework solely requires one ancilla qubit $n_a=1$ for non-unitary embeddings.
Despite differences in ancilla requirements, the success probability derived in Equation~\eqref{eq:succ_prob_avg} matches averaging the probability in Reference~\cite{Nibbi2024}, which shows an alignment with the LCU characteristics, as confirmed here. 

%%%%%%%%%%%%%%%%%%%%%%%%%%%%%%%%%%%%%%%%%%%%%%%%%%%%%%%%%%%%%%%%%%%%
%%  Applications
%%%%%%%%%%%%%%%%%%%%%%%%%%%%%%%%%%%%%%%%%%%%%%%%%%%%%%%%%%%%%%%%%%%%
\section{Applications}
\label{sec:result}
This section demonstrates the capabilities of the presented operator learning protocol across four different applications, covering quantum and engineering problems. 
The learning process is shaped by specifications of the underlying quantum hardware, which in turn determine the final circuit architecture. 
To this end, key considerations within this demonstration include the availability of short- and long-range interaction in the quantum gates and the qubit connectivity defined by the chip architecture. 
First, the Section~\ref{sec:Quantum} benchmarks our framework by learning propagators arising in native quantum problems such as the simulation of quantum many-body systems (cf. Section~\ref{sec:QuantumSimulation}) and quantum chemistry (cf. Section \ref{sec:QuantumChem}).
In addition, both cases cover comparisons to standard Suzuki--Trotter expansions.
Second, engineering applications are specifically addressed in Section~\ref{sec:Engineering}, where we initially discuss the learning of a second-order finite difference approximation (cf. Section~\ref{sec:Laplace}) and then apply the learning protocol to a problem motivated by Computational Fluid Dynamics (CFD) (cf. Section~\ref{sec:panel}). 
Note that the two engineering applications employ amplitude encoding together with the application of the regularization term given in Equation~\eqref{eq:regu}. 

To keep the benchmark comprehensive, we restrict ourselves to constrained layouts, namely one-dimensional LNN and two-dimensional Lattice (2DL) topologies.
Instead of a simple line array topology (LNN), where inner qubits have two direct neighbors, the inner qubits in the 2DL chip have four neighbor interconnections (cf. Figure~\ref{fig:overview}). 
Additionally, the quantum chemistry application (cf. Section~\ref{sec:QuantumChem}) specifically addresses long-range interacting gates.
For clarity, the hardware specifications for each case are defined individually in the corresponding section.
The complexity and the accuracy of the framework are evaluated based on the relative error, cf. Equation~\eqref{eq:relativeErrorScalar}, and the probabilistic characteristics. 
Appendix~\ref{app:specfro} assesses the ratio of the spectral norm error $\|\mathcal{E}(A,c \,
\text{Tr}_{\text{anc}}[P G])\|_2$ to the Frobenius norm error $\|\mathcal{E}(A,c \,
\text{Tr}_{\text{anc}}[P G])\|_\mathrm{fro}$ for the final result of the following applications. 
Furthermore, all simulations have been performed on classical CPU hardware using PyTorch, assuming perfect qubit simulation.

%%%%%%%%%%%%%%%%%%%%%%%%%%%%%%%%%%%%%%%%%%%%%%%%%%%%%%%%%%%%%%%%%%%%
%%  Quantum Applications
%%%%%%%%%%%%%%%%%%%%%%%%%%%%%%%%%%%%%%%%%%%%%%%%%%%%%%%%%%%%%%%%%%%%
\subsection{Quantum Applications}
\label{sec:Quantum}
In this section, we benchmark our algorithm on the compression of the propagator (time-evolution operator), of the time-independent Schrödinger equation,
\begin{equation}
|\psi(t)\rangle = U(t)|\psi(0)\rangle \, ,
\qquad
U(t) = e^{-\mathfrak{j}\mathcal{H}t}\, ,
\end{equation}
where $t \ge 0$ denotes time, $ |\psi(t)\rangle \in \mathbb{C}^{2^n}$ is a quantum many-body state encoded on $n$ qubits, and $\mathcal{H} : \mathbb{C}^{2^n} \rightarrow \mathbb{C}^{2^n}$ is a time-independent many-body Hamiltonian acting on the corresponding Hilbert space. 
Since the Hamiltonian is Hermitian, $\mathcal{H} = \mathcal{H}^{\dag}$, the propagator $U(t)$ is unitary.
Accordingly, the propagator $A\approx U(t)$ is also unitary (st. $A^\dagger A = A A^\dagger = \mathbb{I}$), and, therefore the operator does not require a block-encoding to be applied on quantum hardware. 
As this is valid for all the applications, the success probability stabilizer is $\mu=0$.
These applications are based on second quantization and not on amplitude encoding as they represent quantum many-body systems. 
Interactions are inherently local or decay rapidly with the distance between the qubits.
In this case, we disable the regularizer in the cost function (cf. $\rho=0$ in Equation~\eqref{eq:control_problem_unconstrained}).

Our benchmark problems are chosen to represent two important classes of applications: quantum many-body physics (cf. Section~\ref{sec:QuantumSimulation}) and a quantum chemistry application (cf. Section~\ref{sec:QuantumChem}). 
Although both paradigms are inherently quantum, they exhibit important structural differences.
These include variations in the interaction range, the presence of additional symmetries, and distinct connectivity structures. 
As we discuss below, these differences lead to different requirements and strategies in the optimization procedure.
In both cases, we compare our findings with Trotterized time evolution schemes, often employed to simulate Hamiltonian dynamics on digital quantum hardware.

%%%%%%%%%%%%%%%%%%%%%%%%%%%%%%%%%%%%%%%%%%%%%%%%%%%%%%%%%%%%%%%%%%%%
%%  QUANTUM Simulation
%%%%%%%%%%%%%%%%%%%%%%%%%%%%%%%%%%%%%%%%%%%%%%%%%%%%%%%%%%%%%%%%%%%%
\subsubsection{Quantum Simulation}
\label{sec:QuantumSimulation}
To benchmark our approach in quantum simulations, we consider the 1D transverse-field Ising model (TFIM) with an additional longitudinal field, viz.
\begin{equation}\label{eq:ising}
\mathcal{H} = -g_{zz} \sum_{i=1}^{n-1} {Z}_i {Z}_{i+1} - g_x \sum_{i=1}^{n} {X}_i - g_z \sum_{i=1}^{n} {Z}_i \,,
\end{equation}
where $n$ again denotes the number of qubits. 
The operators ${Z} = |0\rangle\langle 0| - |1\rangle\langle 1|$ and ${X} = |1\rangle\langle 0| + |0\rangle\langle 1|$ correspond to the Pauli-$Z$ and Pauli-$X$ operators acting on the $i$-th qubit, respectively. 
In our numerical experiments, we consider system sizes {$n=4\text{--}12$} and set the coupling parameters to $g_{zz} = 1.0$, $g_x = 0.8$, and $g_z = 0.3$. 
These values are chosen to enable direct benchmarking against \citeauthor{Termanova2024}~\cite{Termanova2024}.
To set the time scale in our simulations, we normalize the Hamiltonian by its spectral norm density $\mathcal{N}={\lVert \mathcal{H}\rVert_2}/{n}\, ,$ and replace $\mathcal{H}$ by $\mathcal{H}/\mathcal{N}$.
This ensures that the rescaled Hamiltonian has unit spectral norm density $\lVert \mathcal{H}\rVert_2/n=1$.

As benchmarks for compressing the propagator $U(t)=e^{-\mathfrak{j}\mathcal{H}t}$,  we employ the first- and second-order Suzuki--Trotter (ST)~\cite{Suzuki1990} expansions
\begin{equation}\label{eq:ST_Ising}
U_{\mathrm{ST}_1}(\delta t)=\prod_{i=1}^{n-1} e^{-\mathfrak{j} h_i \delta t}\,, 
\qquad
U_{\mathrm{ST}_2}(\delta t)=\left(U_{\mathrm{ST}_1}(\delta t/2)\right)^{T} U_{\mathrm{ST}_1}(\delta t/2) \, ,
\end{equation}
where $(\cdot)^T$ denotes the matrix transpose. For a product of matrices, this reverses the ordering of the factors, i.e., $({A}{B})^T = {B}^T {A}^T$, such that $U_{\mathrm{ST}_2}$ corresponds to applying the same sequence of exponentials in reverse order followed by the forward sequence. 
The approximation error satisfies $\varepsilon\!\left(U(\delta t),U_{\mathrm{ST}_m}(\delta t)\right)\propto \delta t^{m+1}$ and the Hamiltonian densities $h_i$ are defined as
\begin{equation}
{h}_i = {f}_i \quad \forall i\in[2,n-2] , \, {h}_1 = {f}_1 - \frac{g_x}{2}{X}_1 - \frac{g_z}{2}{Z}_1\, , \, 
{h}_{n-1} = {f}_{n-1} - \frac{g_x}{2}{X}_n - \frac{g_z}{2}{Z}_n \,,
\end{equation}
where
\begin{equation}
{f}_i = -g_{zz} {Z}_i {Z}_{i+1}
      - \frac{g_x}{2}\left({X}_i + {X}_{i+1}\right)
      - \frac{g_z}{2}\left({Z}_i + {Z}_{i+1}\right)\,.
\end{equation}
We aim to compare trainable quantum circuits with ST expansions over a range of time steps $\delta t$ and different system sizes $n$.
For this reason, we keep the quantum resources close to the Trotter's and pick three different trainable ansätze, viz., 
\begin{equation}
    \begin{gathered}
        G_1 = \prod^{n-1}_{i=1}G_i, \quad \mathcal{G}_i = (i,i+1)\,, \quad 
        G_2 = \prod^{2}_{j=1}\prod^{n-1}_{i=1}G^j_i, \quad \mathcal{G}^j_i = (i,i+1)\,, \\
        G_3 = \prod^{n-2}_{i=1}G_i, \quad \mathcal{G}_i = (i,i+1, i+2)\,.
    \end{gathered}
\end{equation}

The ansatz $G_1$ mirrors the gate count and topology of the first-order expansion $U_{\mathrm{ST}_1}$, cf. Section~\ref{sec:complexity}. 
The ansatz $G_2$ has the same gate count as $U_{\mathrm{ST}_2}$ but different topology. 
Finally, the ansatz $G_3$ is a 3-body generalization of the ansatz $G_1$. 
All three circuits can be efficiently implemented by a 1D LNN hardware layout.
The trainable circuits $(G_1,G_3)$ are always initialized from the first-order expansion $U_{\mathrm{ST}_1}(\delta  t)$.
The circuit $G_2$ is similarly initialized as $\left(U_{\mathrm{ST}_1}(\delta t/2)\right)^2$. 
Across all time steps and system sizes tested, this initialization strategy was found to perform better than initializing the gates as identity operators. 
Especially, for the {$n=12$} instance, we found it significantly more efficient to initialize the optimization from the gates obtained for the optimized {$n=10$} instance, rather than from $U_{\mathrm{ST}_1}$.
Specifically, the two additional qubits are inserted between the qubits at positions $n/2$ and $n/2+1$ of the previously optimized system.
The two, three, or six ($l=2$) additional gates introduced by this enlargement are then initialized from the neighboring optimized gates of the smaller system.

We employ the learning strategy described in Algorithm~\ref{alg:strat}. 
We found that $\texttt{maxit}=10$ steps of individual gate optimization and $\texttt{maxit}\sim10^{3}$ global optimization steps provide adequate performance across all cases considered. 
The learning rate is set to a constant $\eta=0.2$, and the optimization is terminated upon satisfying the convergence criterion $\| \partial\mathcal{J}/\partial G\| \leq \texttt{tol}=10^{-13}$.

Figure~\ref{fig:HamSimDtError} shows the dependence of the relative error $\varepsilon\!\left(U(\delta t),G_i\right)$, cf. Equation~\eqref{eq:relativeErrorScalar}, on the time step $\delta t$, and compares it with that of the ST expansions in Equation~\eqref{eq:ST_Ising} by keeping the system size constant at $n=8$.
The variational circuit \(G_1\) yields an improved error scaling compared to the topologically equivalent first-order Trotter expansion \(U_{\mathrm{ST}_1}\).
Specifically,
${\varepsilon\!\left(U(\delta t),G_{1}\right)}/{\varepsilon\!\left(U(\delta t),U_{\mathrm{ST}_1}\right)}\propto \delta t$, 
indicating an additional suppression by one power of $\delta t$.
For small time steps, the error of $G_1$ is approximately $150$ times smaller than that of $U_{\mathrm{ST}_1}$.
For $G_2$ and $G_3$, the improvement over $U_{\mathrm{ST}_2}$ is roughly three orders of magnitude for small time steps with fitted power-law scalings, $\varepsilon\!\left(U(\delta t),G_{2}\right)\propto \delta t^{4.26}$, $\varepsilon\!\left(U(\delta t),G_{3}\right)\propto \delta t^{5.14}$.

\begin{figure}[t]
\centering
    \vspace{-0.5cm}
    \tikzsetnextfilename{HamSimLegend}
\begin{tikzpicture} 
%\useasboundingbox (-2,-1.29) rectangle (7,4);
    \begin{axis}[
        legend style={draw=black!15!black,legend cell align=center, fill opacity=0.8, draw opacity=1, text opacity=1},
        legend columns=5, 
        hide axis,
        width=0.005\textwidth,
        height=0.005\textwidth,
    ]
        \addlegendimage{red,dotted,mark={x}, mark options={solid}}
        \addlegendentry{\scriptsize$U_{ST_1}$}

        \addlegendimage{red,dotted,mark={o},mark options={solid}}
        \addlegendentry{\scriptsize$U_{ST_2}$}
        
        \addlegendimage{mark={x}}
        \addlegendentry{\scriptsize$G_1$}

        \addlegendimage{mark={o}, mark options={solid}}
        \addlegendentry{\scriptsize$G_2$}

        \addlegendimage{mark={triangle},dashed,mark options={solid}}
        \addlegendentry{\scriptsize$G_3$}
                
        \addplot [draw=none] coordinates {(0,0) (0.001,0.001)};
    \end{axis}
\end{tikzpicture}
    
    \hspace{-1cm}
    \begin{subfigure}[b]{0.45\textwidth}
        \centering
        \caption{\hspace{5.4cm}\vspace{-0.8cm}}
        \newsavebox{\myboxyy}
\savebox{\myboxyy}{%
\tikzexternaldisable
\begin{tikzpicture}
    \begin{loglogaxis}[
        ymin=9e-3, ymax=1,
        xmin=0.8, xmax=1.1,
        scale only axis,
        yminorgrids=true,
        xlabel={ },
        ylabel={ },
    every axis/.append style={line width = 0.5pt},
    grid=major,     
    grid style={line width=.1pt, draw=gray!50},
    width=0.15\textwidth,  
    height=0.18\textwidth,
    xticklabels=\empty,
    ]

    % % Trotter 1 th order
    \addplot[mark=x, color=red, dotted, mark options={solid}] coordinates {
    (0.001, 7.692688492278228e-07) 
    (0.002, 3.0770749105116536e-06)
    (0.005, 1.923169691015213e-05)
    (0.01, 7.692648359173622e-05)   
    (0.02, 0.00030770106702128453)
    (0.05,  0.0019229180050905244)
    (0.1, 0.0076885787781496686)   
    (0.2, 0.030702290164804562)
    (0.5,  0.18894517722484785)
    (1.0,  0.6849100271619925)
     };

   % % Trotter 2 th order
    \addplot[mark=o, color=red, dotted,mark options={solid}] coordinates {
    (0.001, 1.4516712629439617e-10) 
    (0.002, 1.1613365527685594e-09)
    (0.005, 1.8145833669417813e-08)
    (0.01, 1.4516524175261042e-07)
    (0.02, 1.1612762546040391e-06)
    (0.05, 1.8139947956600353e-05)
    (0.1, 0.00014497721530755563)  
    (0.2, 0.0011552994553452054)
    (0.5, 0.01758179788235295)
    (1.0, 0.1293397663034916)
     };
     
    % n =  8 qubits two qubit -  1 layer -- Not covered by termanova -- corresponds to trotter 1st 
     \addplot[mark=x] coordinates {
         (0.01, 5.403814389871029e-07) 
         (0.02, 4.3225957654588925e-06)  
         (0.05, 6.749073078595732e-05) 
         (0.1, 0.0005385043774010231)
         (0.2, 0.004262818483464947)
         (0.5, 0.06195818851901174) % ?
         (1.0, 0.37547966249250275) %?
     };
        % n =  8 qubits two qubit -  2 layer -- Not covered by termanova  - corresponds to trotter 2nd
    \addplot[mark=o,mark options={solid}] coordinates {
        (0.01, 1.3272248016825238e-10) 
        (0.02, 1.0998861620917196e-09)
        (0.05, 3.605117990486911e-08)
        (0.1, 5.188506162220667e-07)
        (0.2, 1.774911906168649e-05) 
        (0.5, 0.001294169987580912) 
        (1.0, 0.03394089653620835) 
        };

     % n =  8 qubits three qubit Termanova R=2
     \addplot[mark=triangle, dashed, mark options={solid}] coordinates {
         (0.01, 4.456279322552675e-12)  
         (0.02, 1.4021388439689136e-10)
         (0.05, 1.368483067086164e-08) 
         (0.1, 4.3705858271144014e-07) 
         (0.2, 1.3876799995604742e-05) 
         (0.5, 0.0012827021515466872) 
         (1.0, 0.12809799552860993)
    };
    \end{loglogaxis}
\end{tikzpicture}
\tikzexternalenable
}
\tikzsetnextfilename{HamSimdtError}
\begin{tikzpicture}
\useasboundingbox (-2,-1.2) rectangle (7,4);
    \begin{loglogaxis}[
    	xlabel={$\delta t$},%  [\  ]},
    	ylabel={$\varepsilon(U(\delta t),*)$},% $\varepsilon$ [\  ]},
        ymin=1e-12, ymax=1,
        xmin=0.01, xmax=1,
        scale only axis,
        width=6cm,
        height=3cm,
         yticklabel style={
            /pgf/number format/fixed,
            /pgf/number format/precision=0,
        },
        yminorgrids=true,
        ytick={1e-12,1e-9,1e-6,1e-3,1e-0},]

    % % Trotter 1 th order
    \addplot[mark=x, color=red, dotted, mark options={solid}] coordinates {
    (0.001, 7.692688492278228e-07) 
    (0.002, 3.0770749105116536e-06)
    (0.005, 1.923169691015213e-05)
    (0.01, 7.692648359173622e-05)   
    (0.02, 0.00030770106702128453)
    (0.05,  0.0019229180050905244)
    (0.1, 0.0076885787781496686)   
    (0.2, 0.030702290164804562)
    (0.5,  0.18894517722484785)
    (1.0,  0.6849100271619925)
     };

   % % Trotter 2 th order
    \addplot[mark=o, color=red, dotted,mark options={solid}] coordinates {
    (0.001, 1.4516712629439617e-10) 
    (0.002, 1.1613365527685594e-09)
    (0.005, 1.8145833669417813e-08)
    (0.01, 1.4516524175261042e-07)
    (0.02, 1.1612762546040391e-06)
    (0.05, 1.8139947956600353e-05)
    (0.1, 0.00014497721530755563)  
    (0.2, 0.0011552994553452054)
    (0.5, 0.01758179788235295)
    (1.0, 0.1293397663034916)
     };
     
    % n =  8 qubits two qubit -  1 layer --corresponds to trotter 1st 
     \addplot[mark=x] coordinates {
         (0.01, 5.403814389871029e-07) 
         (0.02, 4.3225957654588925e-06)  
         (0.05, 6.749073078595732e-05) 
         (0.1, 0.000538504377401025)
         (0.2, 0.004262818483464947)
         (0.5, 0.06195818851901174) % ?
         (1.0, 0.37547966249250275) %?
     };
        % n =  8 qubits two qubit -  2 layer --corresponds to trotter 2nd
    \addplot[mark=o,mark options={solid}] coordinates {
        (0.01, 1.3272248016825238e-10) 
        (0.02, 1.0998861620917196e-09)
        (0.05, 3.605117990486911e-08) %3.27484837068861e-08
        (0.1, 4.971745422914505e-07)
        (0.2, 1.774911906168649e-05) 
        (0.5, 0.001294169987580912) 
        (1.0, 0.03394089653620835) 
        };

     % n =  8 qubits three qubit Termanova R=2
     \addplot[mark=triangle, dashed, mark options={solid}] coordinates {
         (0.01, 4.456279322552675e-12)  
         (0.02, 1.4021388439689136e-10)
         (0.05, 1.368483067086164e-08) 
         (0.1, 4.3705858271144014e-07) 
         (0.2, 1.3876799995604742e-05) 
         (0.5, 0.0012827021515466872) 
         (1.0, 0.12809799552860993)
    };

  \draw [line width=.4pt](axis cs: 0.75,0.008) rectangle (1.0,1.0);
  \draw[-{Stealth[length=3mm]},line width=.4pt]  (0.72,0.0005) -- (axis cs: 0.6,5e-6);
  \draw (axis cs: 0.36, 2e-9) node{\usebox{\myboxyy}};
     
\end{loglogaxis}
\end{tikzpicture}
        \label{fig:HamSimDtError}
    \end{subfigure}
    \begin{subfigure}[b]{0.45\textwidth}
        \centering
        \caption{\hspace{4.8cm}\vspace{-0.8cm}}
        \tikzsetnextfilename{HamSimQubitError}
\begin{tikzpicture}
\useasboundingbox (-2,-1.2) rectangle (7,4);
    \begin{semilogyaxis}[
        xlabel={$n$},% [\  ]},
        ylabel={},
        width=6cm,
        height=3cm,
        xtick={4,6,8,10,12},%,14,16},
        scale only axis,
        ymin=1e-8, ymax=1e-2,
        ]
  
    % % Trotter 1 th order
    \addplot[mark=x, color=red, dotted, mark options={solid}] coordinates {
    (4, 0.005171152543215492) 
    (6, 0.006598252680584546) 
    (8, 0.0076885787781496686) 
    (10, 0.008621547117040874)
    (12, 0.00945483792887039) 
    };

   % % Trotter 2 th order
    \addplot[mark=o, color=red, dotted,mark options={solid}] coordinates {

    (4, 0.00012164930797451605) 
    (6, 0.00013279917680649603) 
    (8, 0.00014497721530755563) 
    (10, 0.00015680929893933446) 
    (12, 0.00016806266494428983)
     };
     
    % n qubits two qubit -  1 layer -- Not covered by termanova -- corresponds to trotter 1st 
    \addplot[mark=x] coordinates {
        (4, 0.0003909598465593581)
        (6, 0.00047378435545690197)
        (8, 0.0005385043774010254)
        (10, 0.000595076480558216) 
        (12, 0.0006463340862071002)
    };
        % n qubits two qubit -  2 layer -- Not covered by termanova  - corresponds to trotter 2nd
    \addplot[mark=o,mark options={solid}] coordinates {
        (4, 2.980882068841486e-07) 
        (6, 4.2311265461737836e-07)
        (8, 4.971745422914505e-07) 
        (10, 8.559601219912465e-07)
        (12, 2.006117445277739e-06)
    };

    % n qubits three qubit Termanova R=2
    \addplot[mark=triangle, dashed, mark options={solid}] coordinates {
        (4, 2.8650180627154586e-07) 
        (6, 3.8348583902915466e-07) 
        (8, 4.3705858267108834e-07) 
        (10, 4.805661157623695e-07) 
        (12, 5.192158296561345e-07)
    };

\end{semilogyaxis}
\end{tikzpicture}
        \label{fig:HamSimQubitError}
    \end{subfigure}
    \vspace{-0.9cm}
    \caption{Application of the learning algorithm to multiple quantum many-body scenarios and comparison against commonly used Trotter decompositions in terms of timestep $\delta t$ with constant system size $n=8$ (\subref{fig:HamSimDtError}), system size $n$ with constant timestep $\delta t=0.1$ (\subref{fig:HamSimQubitError}), and relative error $\varepsilon$. 
    Same symbols denote identical quantum resources.}
    \label{fig:HamSim}
    \vspace{-0.2cm}
\end{figure}

The approximate algebraic error scaling observed for the learned circuits suggests the existence of topologically equivalent Trotter formulas with the same order of accuracy.
Indeed, the structure of the Ising Hamiltonian (cf. Equation~\eqref{eq:ising}) enables the construction of a second-order Trotter formula topologically equivalent to $G_1$. 
This formula is obtained by grouping the operators in the sequence $U_X(\delta t/2) U_Z(\delta t) U_X(\delta t/2)$, where $U_X$ contains the Pauli-$X$ terms and $U_Z$ contains all terms composed of Pauli-$Z$ operators. 
For $G_2$ and $G_3$, however, we did not identify Trotter formulas exhibiting the same algebraic error scaling.

Importantly, these results are not artifacts of fine-tuning on small systems.
Figure~\ref{fig:HamSimQubitError} shows that, at fixed time step $\delta t = 0.1$, the error remains stable across all system sizes, $n=4\text{--}12$.
The only exception is $G_2$ at $n=12$, where the optimization appears to become less effective. 
In general, our results show a remarkable improvement and are consistent with recent related work~\cite{Termanova2024,Gibbs2025}, indicating that the proposed optimization can compress native quantum dynamics beyond conventional product-formula approximations. 

In a direct comparison, the $G_3$ ansatz yields results approximately one order of magnitude better than those reported in Reference~\cite{Termanova2024}. A corresponding comparison for the $G_1$ and $G_2$ ansatz is not possible because of insufficient available data.
Reference~\cite{Termanova2024} additionally presents results for $r=r^j=4,5$-qubit gates, for which similar improvements are expected. 
Furthermore, it is found that the quality of the optimized circuit depends sensitively on the choice of initialization. 
For sufficiently small time steps $\delta t$, ST formulas provide a natural and effective initialization. 
For larger time steps and system sizes, however, this initialization becomes less reliable, and confirms the effect of improved initialization strategies.

This behavior is consistent with the Lieb--Robinson bounds~\cite{Lieb1972} governing information propagation in locally interacting quantum systems.
These bounds imply that information spreads at a finite velocity that remains constant in time. 
As a result, for evolution times $\delta t \sim \mathcal{O}(1)$, the unitary operator $U(\delta t)$ effectively entangles only qubits separated by a distance of order $\mathcal{O}(1)$.
Consequently, the optimal gates are not expected to be significantly affected by the addition of new qubits, provided that the system size remains much larger than this distance.
This confirms the insights found for the initialization strategy employed for jump-starting the larger systems.

%%%%%%%%%%%%%%%%%%%%%%%%%%%%%%%%%%%%%%%%%%%%%%%%%%%%%%%%%%%%%%%%%%%%
%%  QUANTUM CHEMISTRY
%%%%%%%%%%%%%%%%%%%%%%%%%%%%%%%%%%%%%%%%%%%%%%%%%%%%%%%%%%%%%%%%%%%%
 \subsubsection{Quantum Chemistry}
 \label{sec:QuantumChem}
In this section, we consider quantum chemistry models, whose fundamental degrees of freedom are interacting fermionic modes rather than spins. 
To simulate intrinsically fermionic systems on a qubit-based QC, the fermions need to be mapped to qubits. 
Fermionic creation and annihilation operators obey anticommutation relations, mapping these Hamiltonians to qubits requires transformations such as Jordan-Wigner or Bravyi-Kitaev~\cite{Bravyi2002,Seeley2012}, which can introduce parity strings and nonlocal Pauli operators. 
This additional nonlocal structure makes their digital simulation more demanding than the spin models considered in the previous section.

\smallskip
We consider the Pariser--Parr--Pople (PPP) model~\cite{Pariser1953,Pople1953} as a minimal semi-empirical description of the $\pi$-electron sector of butadiene~\cite{Barford2013}, $\mathrm{CH_2=CH-CH=CH_2}$. In this approximation, the active space is restricted to the four carbon $p_z$ orbitals, $N_{\rm orb}=4$, participating in the conjugated $\pi$ system. 
Each spatial orbital supports two spin states, so that the fermionic problem contains $N_{\rm so}=2\cdot N_{\rm orb}=8$ spin orbitals.

The PPP Hamiltonian is
\begin{equation}
\mathcal{H}_{\rm PPP}
=
-\sum^{N_{\rm orb}-1}_{i=1}\mkern-10mu\sum_{\sigma \in \{\uparrow,\downarrow\}}
t_{i}
\left(
c^\dagger_{i\sigma}c_{i+1\sigma}
+
c^\dagger_{i+1\sigma}c_{i\sigma}
\right)
+
U\sum^{N_{\rm orb}}_{i=1} n_{i\uparrow}n_{i\downarrow}
+
\sum^{N_{\rm orb}}_{j=2}\sum^{j-1}_{i=1} V_{ij}(n_i-1)(n_j-1)\, ,
\label{eq:ppp_hamiltonian}
\end{equation}
where $i,j\in\{1,\ldots,N_{\rm orb}\}$ label the carbon sites along the conjugated chain and $\sigma,\tau\in\{\uparrow,\downarrow\}$ label the spin projections.
The fermionic operators obey the canonical anticommutation relations $\{c_{i\sigma},c_{j\tau}^{\dagger}\}=\delta_{ij}\delta_{\sigma\tau}$, with $\{c_{i\sigma},c_{j\tau}\}=\{c_{i\sigma}^{\dagger},c_{j\tau}^{\dagger}\}=0$.
We define the spin-resolved number operator $n_{i\sigma}=c_{i\sigma}^{\dagger}c_{i\sigma}$ and the total site occupation $n_i=n_{i\uparrow}+n_{i\downarrow}$. 

The first term of Equation~\eqref{eq:ppp_hamiltonian} describes hopping between neighboring $p_z$ orbitals, the second term is the on-site Coulomb repulsion, and the third term contains long-range Coulomb repulsion with a neutralizing background corresponding to one $\pi$ electron per carbon atom.
The Hamiltonian conserves the total fermion number $[\mathcal{H}_{\rm PPP},N_f]=0$, where $N_f = \sum^4_i n_{i}$. 
Butadiene is described by the half-filling sector ($N_f = 4$) of the Hamiltonian. 

For the numerical examples below, we adopt the standard PPP parameters for polyenes~\cite{Chandross1997}: a uniform hopping parameter $t_{i}= 2.4 ~{\rm eV}$ and on-site interaction $U = 11.26~{\rm eV}$. 
The long-range Coulomb matrix elements are chosen using the \citeauthor{Ohno1964} interpolation formula~\cite{Ohno1964} as $ V_{ij} = 14.397\, \left({\left( \frac{14.397}{U} \right)^2+ {d_{ij}}^2 }\right)^{-1/2} {\rm eV}$, where $d_{ij}=1.4|i-j|~\text{\AA}$ -- measured in Angstrom (\AA) -- is assumed for a uniform butadiene chain.
We adopt the interleaved spin-orbital ordering
\[
\begin{array}{c|cccccccc}
\text{qubit } p
&1&2&3&4&5&6&7&8\\
\hline
\text{spin orbital}
&1\uparrow&1\downarrow
&2\uparrow&2\downarrow
&3\uparrow&3\downarrow
&4\uparrow&4\downarrow 
\end{array}\,,
\]
to reduce the dimension of each degree of freedom to two and allow a mapping to a spin-qubit model. 
We then map the fermionic operators associated with the eight spin orbitals to eight qubits using the Jordan--Wigner transformation~\cite{Seeley2012}, viz.,
\begin{equation}
c_{i\sigma}
=
\left(
\prod_{\ell=1}^{p(i,\sigma)-1} Z_\ell
\right)
\frac{
X_{p(i,\sigma)}+{\mathfrak{j}}Y_{p(i,\sigma)}
}{2}\,,
\qquad
c_{i\sigma}^{\dagger}
=
\left(
\prod_{\ell=1}^{p(i,\sigma)-1} Z_\ell
\right)
\frac{
X_{p(i,\sigma)}-{\mathfrak{j}}Y_{p(i,\sigma)}
}{2}\,.
\end{equation}

Thus, the qubit PPP Hamiltonian for butadiene has the form
\begin{equation}
\mathcal{H}^{\rm qubit}_{\rm PPP}
=
E_{\rm shift}
+
\sum_{p=1}^{N_{\rm so}} h_p Z_p
+
\sum^{N_{\rm so}}_{q=2}\sum^{q-1}_{p=1} W_{pq} Z_p Z_q
-
\frac{1}{2}
\sum_{p=1}^{N_{\rm so}-2}
t_p
\left(
X_p Z_{p+1} X_{p+2}
+
Y_p Z_{p+1} Y_{p+2}
\right)\,,
\end{equation}
where $E_{\rm shift} = U$ is a constant energy offset, $t_p = t_i = 2.4 ~{\rm eV}$, $h_q = -U/4$ and
\begin{equation}
W_{pq}
=
\begin{cases}
\dfrac{U}{4},
& p=(i,\uparrow),\; q=(i,\downarrow)\,,
\\[8pt]
\dfrac{V_{ij}}{4},
& p=(i,\sigma),\; q=(j,\tau),\; i\neq j\,.
\end{cases}
\end{equation}

As a consequence of the interleaved ordering, the mapped Hamiltonian is at most three-local in the Pauli representation: The Coulomb sector is diagonal and two-local, while the hopping sector is three-local. 
This locality structure is important for digital simulation schemes in which native gates may act on non-neighboring qubits but are restricted to few-body interactions.

Due to the complexity of the PPP Hamiltonian, we perform only a first-order ST expansion. 
The decomposition is organized to reduce the number of elementary evolution blocks by combining terms that can be implemented on the same few-qubit support.
In particular, the three-body hopping contributions are grouped with local and interaction terms that act on the same contiguous support, while the remaining long-range density-density interactions are implemented as separate two-body gates.
The resulting first-order step is written as $U_{\mathrm{ST}_1}(\delta t) = U^3_{\mathrm{ST}_1}(\delta t) U^2_{\mathrm{ST}_1}(\delta t) $ and consists of 21 gates in total.
The contribution $U^3_{\mathrm{ST}_1}$ contains six three-body gates associated with the hopping terms, and $U^2_{\mathrm{ST}_1}$ contains 15 two-body gates corresponding to the remaining density-density interactions, with the gate layout specified below.

\begin{figure}[t]
    \centering
    \hspace{0.2cm}
    \begin{subfigure}[t]{0.48\textwidth}
        \caption{\hspace{7cm}\vspace{-0.8cm}}
        \centering
        \vspace{1.1cm}
        \includegraphics[width=\textwidth]{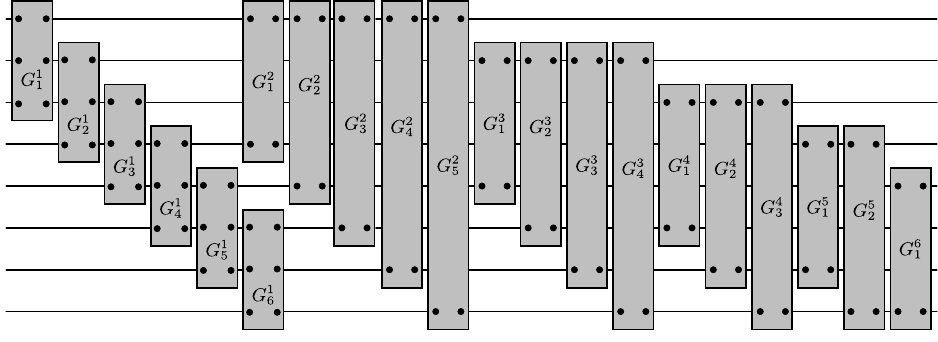}
        \label{fig:qchemcirc}
    \end{subfigure}
    \hfill
    \begin{subfigure}[t]{0.48\textwidth}
        \centering
        \caption{\hspace{6.7cm}\vspace{-0.8cm}}
        \tikzsetnextfilename{QChemdterr}
\begin{tikzpicture}
\useasboundingbox (-2,-1.2) rectangle (7,4);
    \begin{loglogaxis}[
    	xlabel={$\delta t$},% [\  ]},
        ylabel={$\varepsilon(U(\delta t),*)$},% $\varepsilon$ [\  ]},
        ymin=1e-3, ymax=2,
        xmin=0.01, xmax=1,
        legend style={draw=black!15!black,legend cell align=center, fill opacity=0.8, draw opacity=1, text opacity=1, at={(0.98,0.23)}},
        scale only axis,
         yticklabel style={
            /pgf/number format/fixed,
            /pgf/number format/precision=0,
        },
        legend columns=3,
        %yminorgrids=true,
        ytick={1e-3,1e-2,1e-1,1e+0,1e1},
        ]

 % % U Comp-U_exact 
    %||U_comb - U_exact||_F/  ||U_exact||_F    values  
    \addplot[mark=x, color=red, dotted, mark options={solid}] coordinates {
    (0.01, 0.0024351018338481874)   
    (0.02, 0.009724433986369761) 
    (0.05, 0.06006712232783729)
    (0.1, 0.2295878835541196)   
    (0.2, 0.740924361721625)
    (0.5, 1.3540448156296332)
    (1.0,  1.44411755366841)
     };
    \addlegendentry{\scriptsize $U_\mathrm{ST}$}

    % n =  8 qubits 1 layer 3 qubits + 3 layers two qubit (long range) 
     \addplot[mark=x] coordinates {
          (0.01, 0.001990095513677874) %U2 and U3 opt  0.001981937005365251
          (0.02, 0.007931393602457548) %U2 and U3 opt   0.007804072561877454
          (0.05, 0.048514941108041554) %U2 and U3 opt  0.04434125950755523
          (0.1, 0.1808477940278423) %U2 and U3 opt   0.14137912293402505
          (0.2, 0.5798726267647817)  %U2 and U3 opt  0.4055546512408584
          (0.5, 0.9471690381828755) %U2 and U3 opt   
          (1.0, 1.247943698870871) %U2 and U3 opt   0.8606481405334323
     };

    \addlegendentry{\scriptsize $G$}

    %DIFFERENCES in percent --  re,lative 
     \addplot[mark=o, orange] coordinates {
         (0.01, 0.22361053382201823) %U2 and U3 opt 
          (0.02, 0.2260687684641749) %U2 and U3 opt  
          (0.05, 0.23811594852953277) %U2 and U3 opt  
          (0.1,  0.2695088971821993) %U2 and U3 opt  
         (0.2,  0.2777363985180348) %U2 and U3 opt  
         (0.5, 0.42957039455949775) %U2 and U3 opt  
          (1.0,  0.1571976804522799) %U2 and U3 opt  
     };

    \addlegendentry{\scriptsize $\frac{\varepsilon_\mathrm{ST}}{\varepsilon_\mathrm{G}}-1$}

    % % n =  8 qubits 2 layer 3 qubit gates
    %  \addplot[mark=x, dashed, mark options={solid}] coordinates {
    %      (0.01, 0.01) 
    %       (0.02, 0.01)
    %       (0.05, 0.01)
    %       (0.1,0.01)
    %      (0.2, 0.01)
    %      (0.5, 0.01)
    %       (1.0, 0.01)
    %  };

    % \addlegendentry{$\varepsilon_{O_3}$ $n_q=3,\, d=2$}

 % %DIFFERENCES in percent --  re,lative 
 %     \addplot[mark=o, orange,dashed, mark options={solid}] coordinates {
 %         (0.01, 0.1) %U2 and U3 opt 
 %          (0.02,  0.1)
 %          (0.05,  0.1)
 %          (0.1,   0.1)
 %         (0.5,  0.1)
 %          (1.0, 0.1)
 %     };

 %    \addlegendentry{$1-\varepsilon_T/\varepsilon_{O_3}$}

\end{loglogaxis}
\end{tikzpicture}
        \label{fig:qchemerror}
    \end{subfigure}
    \vspace{-0.7cm}
    \caption{(\subref{fig:qchemcirc}) Quantum circuit implementation of the butadiene PPP Hamiltonian, consisting of $21$ nearest-neighbor and long-range interaction gates.
(\subref{fig:qchemerror}) Time-step-induced errors as a function of the time-step size for the ST and learned-circuit propagators.}
    \label{fig:qchem}
    \vspace{-0.2cm}
\end{figure}

We solve the problem described above, keeping the ST resources constant and the system aligned to $n = N_{\rm so} = 8$. 
For the learning, we employ $\texttt{maxit}=10$ steps of individual gate optimization and $\texttt{maxit}\sim10^3$ global optimization steps.
The learning rate has been set to $\eta=10^{-5}$ for short-range and $\eta=10^{-3}$ for long-range interaction and 
the gradient criterion is set as $\| \partial\mathcal{J}/\partial G\| \leq \texttt{tol}=10^{-15}$.
This short- and long-range interacting ansatz,  illustrated for $n=8$ in Figure~\ref{fig:qchemcirc}, is given by
\begin{equation}
    \begin{gathered}
                G = G_2 G_1,\, \quad \text{ where } \\
                G_1 = \prod^{1}_{j=1}\prod^{n-2}_{i=1}G^j_i, \quad \mathcal{G}^j_i = (i,i+1,i+2)\,, \quad 
                G_2 = \prod^{n-3}_{j=1}\prod^{n-2-j}_{i=1}G^j_i, \quad \mathcal{G}^j_i = (j,i+j+2)\,.
    \end{gathered}
\end{equation}

Figure~\ref{fig:qchemerror} compares the error of the optimized variational circuit, $\varepsilon(U(\delta t),G)$, with that of the first-order ST approximation, $\varepsilon(U(\delta t),U_{\mathrm{ST}_1}(\delta t))$, for several time steps $\delta t$.
We initialize the learning procedure from the first-order ST expansion. 
We observe a consistent reduction in the error of approximately ${\varepsilon_{\mathrm{ST}}}/{\varepsilon_\mathrm{G}}-1\simeq 20\%$ across the full range of time steps considered. 

This improvement is appreciable, although substantially smaller than the orders-of-magnitude reduction obtained for the Ising Hamiltonian in the previous section. 
We attribute this difference to the lower sparsity of the Trotterized gate generators in the PPP model, which restricts the ability of the learned circuit to improve significantly over the first-order ST initialization. 
We also tested extensions of the ansatz, including an additional three-body gate layer and extra two-body gates, but did not observe a significant improvement over the results reported here.

%%%%%%%%%%%%%%%%%%%%%%%%%%%%%%%%%%%%%%%%%%%%%%%%%%%%%%%%%%%%%%%%%%%%
%%  Engineering Applications
%%%%%%%%%%%%%%%%%%%%%%%%%%%%%%%%%%%%%%%%%%%%%%%%%%%%%%%%%%%%%%%%%%%%
\subsection{Engineering Applications}
\label{sec:Engineering}
In this section, we explore the application of the operator learning protocol to engineering problems. 
The first case concerns the accurate programming of the finite-difference operator associated with the centered second-order approximation of second derivatives, cf. Section~\ref{sec:Laplace}.
This application is broadly relevant across computational science and engineering, as it underlies the solution of partial differential equations on quantum hardware.

Secondly, in Section~\ref{sec:panel}, we focus on learning a dense operator that arises in CFD during the analysis of inviscid flow around a two-dimensional airfoil, modeled using the Hess-Smith panel method~\cite{Hess1967}. 
Both applications employ amplitude encoding, whereby the value of the classical field at each grid point is encoded in the amplitude of the quantum state associated with the corresponding grid-point basis state.
The operators are encoded using the \textit{big-endian} notation.
In this setting, local features of the classical field may result in long-range quantum correlations. 
Consequently, reducing local errors in the classical field often involves fine-tuning multiple local gates, making such errors challenging to correct.
This provides a useful benchmark for the $\ell_2$-regularization term in Equation~\eqref{eq:regu}, controlled by $\rho$, which is actively used in both sections.
Furthermore, the success probability stabilizer $\mu$ is chosen independently for each experiment. 

Unlike the quantum mechanics applications discussed in Section~\ref{sec:Quantum}, the matrices encountered here are generally non-unitary and thus necessitate block encoding. Throughout the experiments, we assess the framework’s complexity and accuracy by analyzing relative errors, success probabilities, and computational effort. 

%%%%%%%%%%%%%%%%%%%%%%%%%%%%%%%%%%%%%%%%%%%%%%%%%%%%%%%%%%%%%%%%%%%%
%%  simple Laplacian 
%%%%%%%%%%%%%%%%%%%%%%%%%%%%%%%%%%%%%%%%%%%%%%%%%%%%%%%%%%%%%%%%%%%%
\subsubsection{Finite Approximation of Second Derivative}
\label{sec:Laplace}
In the following, we consider the prototypical problem of implementing the discretized second spatial derivative, represented by the one-dimensional Laplace operator. 
Without loss of generality, we assume a spatial equidistant one-dimensional finite difference discretization, i.e., $x_{i+1}-x_{i} = \Delta x$, with $2^{n_\text{s}}$ supporting points $i$. 
For the approximation of the second spatial derivative of some generic scalar field $f(x)$, we employ a second-order accurate central difference scheme, viz.,
\begin{equation}
    \frac{\partial^2 f}{\partial x^2} \approx \frac{f_{i-1}-2f_i+f_{i+1}}{\Delta x^2}\, .
    \label{eq:laplace_disc}
\end{equation}
Furthermore, we assume homogeneous Dirichlet boundary conditions for the scalar field $f(x)$, which compiles the domain with non-cycling but closed boundary conditions. 
Rewriting Equation~\eqref{eq:laplace_disc} in matrix-vector form and including the boundary conditions, results in the matrix $A \in \mathbb{R}^{2^{n_\text{s}}\times 2^{n_\text{s}}} \,$ defined as 
\begin{equation}
\label{eq:derivative}
    {A} = \begin{bmatrix}
        -2 & 1 & 0 & \dots & 0 \\
        1 & -2 & 1&\ddots     \\
        && \ddots \\
        &\ddots & 1 & -2 & 1 \\
        0 & \dots & 0 & 1 & -2 \\
    \end{bmatrix} \, .
\end{equation}
Exploiting the tridiagonal Toeplitz structure, the operator ${A}$ is efficiently represented by TTs with a resource scaling of $\mathcal{O}(n_\text{s})$.
In particular, it can be represented as a TT of $n_\text{s}$ local tensors with auxiliary dimension $R({A})=5$ for any grid size \cite{Kazeev2012}. 
\citeauthor{Termanova2024}~\cite{Termanova2024} applies this finding \cite{Kazeev2012} for the initialization of their tensor quantum programming routine. 
In contrast, our work goes beyond this low-bond decomposition by leveraging correlations over all scales. 
Accordingly, it should be understood as an alternative learning approach (cf. Section~\ref{sec:intro}).
We note that the operator in Equation~\eqref{eq:derivative} cannot be written as a product or a sum of operators with local (range-$\mathcal{O}(1)$) operator densities. 
As a result, the exact encoding of the operator requires multi-body gates, and finding an accurate approximation by limiting the gate size is therefore challenging. 

The learned operator is assessed by the relative error and the success probability. 
Additionally, in this application, the system size and the qubit connectivity are varied. 
The base gate size $r^j$ is estimated by assuming an auxiliary dimension of $R=5$~\cite{Kazeev2012}.
Therefore, we employ four-qubit gates with a gate size $r = r^j = \lceil \log_2(R)\rceil + 1 = 4$, as in unitary tensor-network decompositions~\cite{Termanova2024}.
The number of gates per layer is set to $m^j = m = n-3$, which yields overlapping gates.
The number of layers has been chosen heuristically as $l=4$, since additional layers did not reduce the cost function further.
As the operator $A$ is not unitary, we use $n_\text{a}=1$ ancilla qubit for block encoding throughout.
Furthermore, we employ a constant ansatz emulating an LNN-fashioned topology, given by the gate layout
\begin{equation}
    G_\text{LNN} = \prod^{4}_{j=1}\prod^{n-3}_{i=1}G^j_i, \quad \mathcal{G}^j_i = (i,i+1,i+2,i+3)\,.
    \label{eq:G_LNN}
\end{equation} 
We also employ a variant of 2DL topology for which the ansatz is inspired by a block encoding using quantum signal processing.
Every gate in quantum signal processing (local rotation and signals) is applied to the ancilla qubit, a feature we incorporated in our ansatz,
\begin{equation}
    G_\text{2DL} = \prod^{4}_{j=1}\prod^{n-3}_{i=1}G^j_i,  \quad \mathcal{G}^j_i = (1,i+1,i+2,i+3)\,.
        \label{eq:G_2DL}
\end{equation}
This ansatz particularly addresses a 2D lattice chip topology, with qubit $1$ located at the center and the remaining qubits arranged peripherally.
For comparison, both ansätze use only nearest-neighbor four-qubit gates while enabling correlations to propagate over larger length scales across the 2D lattice. 
If the system size grows beyond ten qubits, the realization of $G_\text{2DL}$ requires long-range programmable hardware, since all gates involve the same qubit in position $1$. 
In the following, we assess their accuracy and probabilistic performance, and subsequently demonstrate the hierarchical training strategy together with the impact of regularization using the LNN ansatz.
\bigskip

\begin{figure}[t]
\centering
    \tikzsetnextfilename{LaplaceLegend}
%\useasboundingbox (-2,-1.2) rectangle (7,4);
%\tikzset{external/export next=false}
\begin{tikzpicture} 
    \begin{axis}[
        legend style={draw=black!15!black,legend cell align=center, fill opacity=0.8, draw opacity=1, text opacity=1},
        legend columns=9, 
        hide axis,
        width=0.005\textwidth,
        height=0.005\textwidth,
    ]
        % \addlegendimage{red,dotted,mark={x}, mark options={solid}}
        % \addlegendentry{1st ord. Trotter}
        
        \addlegendimage{mark={x}}
        \addlegendentry{LNN}

        \addlegendimage{mark={o},orange}
        \addlegendentry{2DL}

        % \addlegendimage{color=red, mark={triangle},dashed,mark options={solid}}
        % \addlegendentry{\scriptsize $\varepsilon_\mathrm{LNN}/\varepsilon_\mathrm{2DL}-1$}

        % \addlegendimage{red,dotted,mark={o},mark options={solid}}
        % \addlegendentry{2nd ord. Trotter}

        \addlegendimage{color=red, dotted, mark=square,mark options={solid}}
        \addlegendentry{Operator intrinsics}
        % \addlegendimage{thin,black,mark={triangle},only marks}
        % \addlegendentry{\small CDS$^\text{FD}\;\;$}
        % \addlegendimage{semithick,dashed}
        % \addlegendentry{\small CDS$^\text{VQA}\;\;$}

        % \addlegendimage{thin,black,mark={square},only marks}
        % \addlegendentry{\small QUICK$^\text{FD}\;\;$}
        % \addlegendimage{semithick,dashdotted}
        % \addlegendentry{\small QUICK$^\text{VQA}\;\;$}

        \addplot [draw=none] coordinates {(0,0) (0.1,0.1)};
    \end{axis}
\end{tikzpicture}
    \\
    \vspace{-0.5cm}
    \begin{subfigure}[t]{0.48\textwidth}
       \centering
        \caption{\hspace{6.7cm}\vspace{-0.8cm}}
        \tikzsetnextfilename{LaplaceQubitError}
\begin{tikzpicture}
\useasboundingbox (-1.8,-1.2) rectangle (7,4);
\begin{semilogyaxis}[
        xlabel={$n_\text{s}$},% [\  ]},
    	ylabel={$\varepsilon(A,c\,\text{Tr}_{\text{anc}}[P G])$},% 
        width=6cm,
        height=3cm,
        ymax=1,
        ymin=10^-15,
        ytick={10^-1,10^-3,10^-6,10^-9,10^-12,10^-15},
        xtick={4,5,6,7,8,9,10},
        ]

    % LNN  
    \addplot[mark=x] coordinates {
        (4, 6.590633703257457e-14)
        (5, 2.6345379381783247e-13)  
    	(6, 0.003887231067374235) 
        (7, 0.016581588208131794) 
        (8, 0.039954993657100764)%
        (9, 0.08516268350914162)
    };

      % 2D  
    \addplot[mark=o,orange] coordinates {
        (4, 6.451409168218348e-14) 
        (5, 4.212466862573425e-13)  
    	(6, 0.0018844564059488748) 
        (7, 0.004945203379089959)  
    	(8, 0.005190421393088463)
    	(9, 0.016718277638638687)
    };

\end{semilogyaxis}
\end{tikzpicture}
        \label{fig:LaplaceQubitError}
    \end{subfigure}
        \hspace{0.1cm}
    \begin{subfigure}[t]{0.48\textwidth}
        \centering
        \caption{\hspace{6.7cm}\vspace{-0.8cm}}
        % This file was created with tikzplotlib v0.10.1.
\tikzsetnextfilename{Laplace_SuccProb}
\begin{tikzpicture}
\useasboundingbox (-1.8,-1) rectangle (7,4);
\begin{axis}[
legend style={fill opacity=0.8, draw opacity=1, text opacity=1, at={(0.9,0.8)}},
tick pos=left,
ymin=0.2, ymax=0.4,
xlabel={$n_\text{s}$}, %[\  ]},
ylabel={$p$},%[\  ]},
scaled ticks=false,
xtick={4,5,6,7,8,9,10},
y tick label style={
text width=2.5em
},
width=6cm,
height=3cm,
%ymin=0.001, ymax=1.01,
%ymode=log
]

%Operator intrinsic success probability red dashed
\addplot[color=red, mark=square, dotted, mark options={solid}] 
table[col sep = comma , row sep=\\]{%
4, 0.37352032 \\
5, 0.37277982\\
6, 0.37348289\\
7, 0.37413437\\
8, 0.3745397\\
9, 0.37476289 \\
%10, 0.37487969\\    
     };
     
%LNN
\addplot [mark=x]
table[col sep = comma , row sep=\\]{%
4, 0.3684139588996937\\
5, 0.37277845280008126\\ 
6, 0.22952963121296344\\
7, 0.24288728307009436\\  
8, 0.3541383859582413\\
9, 0.3272516440274525\\
};

%2DL
\addplot [mark=o,orange]
table[col sep = comma , row sep=\\]{%
4, 0.3725495279170323\\
5, 0.3644138638174338\\ 
6, 0.33897785349612913\\
7, 0.2275081388301963\\  
8, 0.23178181812358017\\
9, 0.2817164680655777\\
};

\end{axis}
\end{tikzpicture}
        \label{fig:LaplaceSucProb}
    \end{subfigure}
    \\
    \vspace{-0.7cm}
    \begin{subfigure}[t]{0.48\textwidth}
            \centering
                \caption{\hspace{6.6cm}\quad}
                \hspace{0.8cm}
                \includegraphics[width=0.8\textwidth]{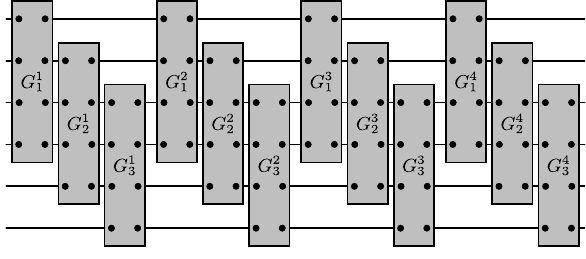}
                \label{fig:LaplaceLNN}
    \end{subfigure}
    \hspace{0.1cm}
      \begin{subfigure}[t]{0.48\textwidth}
            \centering
                \caption{\hspace{6.6cm}\quad}             
                \hspace{0.8cm}
                \includegraphics[width=0.8\textwidth]{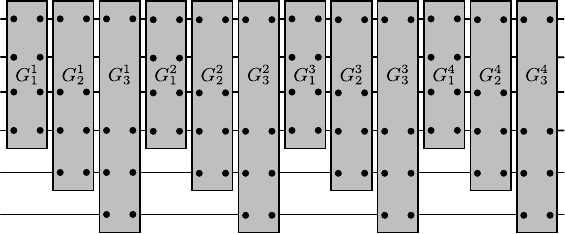}
                \label{fig:Laplace2DL}
    \end{subfigure}
    \caption{Operator learning of the discrete second derivative operator for varying system sizes $n_\text{s}$. The relative error $\varepsilon(A,c\operatorname{Tr}_{\mathrm{anc}}[P G])$ and success probability $p$ are shown in (\subref{fig:LaplaceQubitError}) and (\subref{fig:LaplaceSucProb}), respectively. 
    The LNN and 2DL learnable circuit architectures used for $n=6$ qubits are shown in (\subref{fig:LaplaceLNN}) and (\subref{fig:Laplace2DL}).}
    \label{fig:Laplace}
\end{figure}

%%%%%%%%%%%%%%%%%%%%%%%%%%%%%%%%%%%%%%%%%%%%%%%%%%%%%%%%%%%%%%%%%%%%
\paragraph{Success Probability and Accuracy} are first evaluated to characterize their scaling behavior with increasing system size $n_\text{s}$.
The maximum number of iterations is set to $\texttt{maxit}\sim10^{3}$, with the final training stage extended to $\texttt{maxit}\sim10^{4}$ iterations.
Throughout all experiments, the learning rate is kept constant at $\eta = 0.02$ and regularization is disabled ($\rho=0$).
The weight of the stabilizing term $\mu$ is adapted to the attainable accuracy, i.e., $\mu=10^{-18}\,(n_\text{s}=4)$,
$10^{-15}\,(n_\text{s}=5)$, and $10^{-5}\,(n_\text{s}=6\text{--}9)$,
such that it prevents $c \gg c_\text{ideal}$ without limiting the accuracy,
cf. Equation~\eqref{eq:c_sol} and Appendix~\ref{app:influence_c}.
In addition to the iteration limit, convergence is determined by the gradient criterion $\| \partial\mathcal{J}/\partial G\|\leq\texttt{tol}=10^{-9}$.
To assess the sensitivity of the learning protocol to the underlying hardware connectivity -- one-dimensional (1D) versus two-dimensional (2D) chip topologies -- the experiments compare the two ansätze~\eqref{eq:G_LNN} and \eqref{eq:G_2DL}.
The corresponding results are presented in Figure~\ref{fig:Laplace}, where Subfigures~(\subref{fig:LaplaceLNN}) and (\subref{fig:Laplace2DL}) illustrate representative quantum circuit implementations for $n=6$ qubits.
We observe a highly accurate representation of the operator for both ansätze for system sizes $n_\text{s} < 6$ qubits, where crosses and circles denote the LNN and 2DL results, respectively.
For both, the accuracy reaches machine precision, while the probability of success approximately converges to the operator’s intrinsic probability of \(p(A/\|A\|_2) \approx 0.3734\), cf. Equation~\eqref{eq:succ_prob_avg}. 
As the system size increases, performance limitations emerge for both ansätze. 
The error increases to values on the order of $10^{-2}$, yet remains approximately constant with further growth of the system size. 
Likewise, the success probability $p$ exhibits only moderate variations and remains within the same order of magnitude.
Overall, the results demonstrate improvements compared to literature \cite{Termanova2024}, particularly visible for smaller system sizes. The enhancements are reflected not only in the reconstruction error (cf. Equation~\eqref{eq:error_rel}) but also, and more notably, in the achieved success probabilities.

Training the discretized second derivative operator $A$ requires the representation of long-range interactions.
For smaller system sizes, these interactions can be captured more accurately using a fixed arrangement of $4$-qubit gates. 
As the system size increases, the gates span progressively smaller portions of the quantum register, thereby reducing the effective expressiveness of the ansatz. 
Together with potential BP issues during optimization, arising from the absence of an accurate initialization protocol, this provides a plausible explanation for the observed behavior.
Comparing the two architectures, the 2DL ansatz outperforms the LNN approach. 
The differences can be attributed to the ansatz's ability to realize long-range interactions more effectively by coupling each gate to the first register qubit, i.e., the ancilla.

\bigskip
\paragraph{Training Strategy and Regularization} were introduced in Section~\ref{sec:opt} and are now evaluated quantitatively. 
In the context of amplitude encoding, the ordering and structure of the gate sequence during optimization are expected to strongly influence the learned representation. 
Since the \(n_\text{s}=5\) qubit case already exhibits excellent agreement, with errors approaching machine precision (cf.~Figure~\ref{fig:Laplace}), the following analysis focuses on the more challenging \(n_\text{s}=6\) qubit case. 

\smallskip
To assess the influence of the optimization strategy, the matrix \(A\) for a system size of \(2^6\) is trained under three different configurations: A) without hierarchical optimization, B) with hierarchical optimization, and C) with hierarchical optimization and additional $\ell_2$-regularization.
 
In experiment A, the hierarchical optimization is disabled, such that only a global optimization (cf.~Algorithm~\ref{alg:strat} -- line~\ref{alg:strat:op3}) is performed.
In the other two experiments, B~and~C, the hierarchical optimization stages are performed analogously, each with \(\texttt{maxit}\sim10^3\) iterations. 
For experiment C, a regularization term with weight \(\rho = 4\!\times\! 10^{-8}\) is included, yielding an additional contribution to the objective function. 
For all configurations, the final optimization stage is executed with \(\texttt{maxit}\sim10^4\) iterations, and the convergence criterion on the derivative's norm is set to $\| \partial\mathcal{J}/\partial G\|
\leq\texttt{tol}=10^{-7}$.
All experiments employ the same numerical stability parameter for the success probabilities, viz. $\mu=10^{-6}$, as well as an identical initial learning rate of \(\eta = 0.2\), which is adaptively modified within the L-BFGS optimization procedure.
Compared with the system-size study above (cf.~Figure~\ref{fig:Laplace}), the stabilizing penalty is relaxed from $\mu=10^{-5}$ to $\mu = 10^{-6}$ to allow for a further gain in accuracy.
The corresponding results are summarized in Figure~\ref{fig:LaplaceResults}.
\begin{figure}[t]
    \centering
    \vspace{-0.2cm}
    \hspace{-1cm}
    \begin{subfigure}[t]{0.3\textwidth}
        \centering
        \caption{\hspace{4cm}\vspace{-0.5cm}}
        \tikzsetnextfilename{RefMatDirn6}
\begin{tikzpicture}
\useasboundingbox (-1.5,-1) rectangle (5,4);
        \begin{axis}[
            %view={0}{90},   % not needed for `matrix plot*' variant
            xlabel={$j$},      
            width=3cm,
            height=3cm,
            xtick pos=upper,
            xticklabel pos=upper,
            ylabel={$i$},      
            grid=none,
            y dir=reverse,
            colorbar,
            xticklabel style = {font=\scriptsize},
            yticklabel style = {font=\scriptsize,xshift=3ex},
            colorbar horizontal,
            ytick={0,31,63},
            xtick={0,31,63},
            xticklabels={1,32,64},
            yticklabels={1,32,64},
            colorbar style={
                %title=$^\circ \mathrm{C}$,
                at={(0,-0.1)},
                anchor=north west,
                font =\scriptsize,
                % yticklabel style={
                %     /pgf/number format/.cd,
                %     fixed,
                %     precision=1,
                %     fixed zerofill,
                %     },
                    height=0.1cm,
            },
            %
            % added these key-values
            enlargelimits=false,
            axis on top,
            point meta min=-2.0,
            point meta max=1.0,
            %
            % uncomment me to show, that there are really no cells plotted
            % when using the second variant (`surf, shader=flat corner')
            %xmax=5,
            %ymax=5,
        ]
            %% gives the mean temperature of the neighbouring cells (4x4 matrix)
            %\addplot3 [surf] file {temp.dat};

            %% gives the temperature of each point where the reference point
            %% is the lower left corner of the cell (also a 4x4 matrix)
            %% (I already asked the package author if this is intended or a bug)
            %\addplot3 [surf,shader=flat corner] file {temp.dat};

            % this should give the desired output (5x5 matrix)
            \addplot [matrix plot*,point meta=explicit] file {figures/data/RefMatDir6q.dat};
        \end{axis}
    \end{tikzpicture}
        \label{fig:LaplaceResultsReference}
    \end{subfigure}
    \begin{subfigure}[t]{0.3\textwidth}
        \centering
        \caption{\hspace{4cm}\vspace{-0.5cm}}
        \label{fig:LaplaceResultsLearned}
        \tikzsetnextfilename{LearMatDirn6}
\begin{tikzpicture}
\useasboundingbox (-1.5,-1) rectangle (5,4);
 \begin{axis}[
            view={0}{90},   % not needed for `matrix plot*' variant
            xlabel={$j$},      
            xtick pos=upper,
            ymin=-0.8, ymax=127.5,
            xmin=-0.7, xmax=127.5,
            width=3cm,
            height=3cm,
            xticklabel pos=upper,
            xtick style={draw=none},     
            ytick style={draw=none},      
            xticklabel pos=upper,
            ylabel={$i$},
            trim axis left,
            baseline,
            trim axis right,
            scale only axis,
            ytick={0,31,63,127},
            xtick={0,31,63,127},
            xticklabels={1,32,64,128},
            yticklabels={1,32,64,128},
            grid=none,
            y dir=reverse,
            colorbar,
            xticklabel style = {font=\scriptsize},
            yticklabel style = {font=\scriptsize,xshift=2ex},
            colorbar horizontal,
            colorbar style={
                %title=$^\circ \mathrm{C}$,
                at={(0,-0.1)},
                anchor=north west,
                font =\scriptsize,
                % yticklabel style={
                %     %/pgf/number format/.cd,
                %     %fixed,
                %     %precision=1,
                %     fixed zerofill,
                %     },
                    height=0.1cm,
            },
            %
            % added these key-values
            enlargelimits=false,
            axis on top,
            point meta min=-2.00,
            point meta max=1.00,
            %
            % uncomment me to show, that there are really no cells plotted
            % when using the second variant (`surf, shader=flat corner')
            %xmax=5,
            %ymax=5,
        ]
            %% gives the mean temperature of the neighbouring cells (4x4 matrix)
            %\addplot3 [surf] file {temp.dat};

            %% gives the temperature of each point where the reference point
            %% is the lower left corner of the cell (also a 4x4 matrix)
            %% (I already asked the package author if this is intended or a bug)
            %\addplot3 [surf,shader=flat corner] file {temp.dat};

            % this should give the desired output (5x5 matrix)
            \addplot [matrix plot*,point meta=explicit] file {figures/data/LearMatDir6q.dat};
        \end{axis}
    \end{tikzpicture}
    \end{subfigure}
    \begin{subfigure}[t]{0.3\textwidth}
        \centering
        \caption{\hspace{4cm}\vspace{-0.5cm}}
        \input{figures/Laplace_Cost}
        \label{fig:LaplaceResultsCost}
    \end{subfigure}
    \\
    \vspace{-0.3cm}
    \hspace{-1cm}
    \begin{subfigure}[t]{0.3\textwidth}
        \centering
        \caption{\hspace{4cm}\vspace{-0.5cm}}
        \label{fig:LaplaceResultsErrorNoHier}
        \tikzsetnextfilename{ErrMatDir_nohier_n6}
\begin{tikzpicture}
\useasboundingbox (-1.5,-1) rectangle (5,4);
        \begin{axis}[
            view={0}{90},   % not needed for `matrix plot*' variant
           ymin=-0.7, ymax=63.4,
            xmin=-0.7, xmax=63.4,
            xtick pos=upper,
            xticklabel pos=upper,
            xlabel={$j$}, 
            width=3cm,
            height=3cm,
            ylabel={$i$}, 
            grid=none,
            y dir=reverse,
            colorbar,
            ytick={0,31,63},
            xtick={0,31,63},
            xticklabels={1,32,64},
            yticklabels={1,32,64},
            xticklabel style = {font=\scriptsize},
            yticklabel style = {font=\scriptsize,xshift=3ex},
            colorbar horizontal,
            colorbar style={
                %title=$^\circ \mathrm{C}$,
                at={(0,-0.1)},
                anchor=north west,
                font =\scriptsize,
                xtick={0.0,0.05,0.1},                
                % yticklabel style={
                %      /pgf/number format/.cd,
                %      fixed,
                %      precision=2,
                %      fixed zerofill,
                %      },
                scaled ticks=false,
                height=0.1cm,
            },
            %
            % added these key-values
            enlargelimits=false,
            axis on top,
            point meta min=0.00,
            point meta max=0.1,
            %
            % uncomment me to show, that there are really no cells plotted
            % when using the second variant (`surf, shader=flat corner')
            %xmax=5,
            %ymax=5,
        ]
            %% gives the mean temperature of the neighbouring cells (4x4 matrix)
            %\addplot3 [surf] file {temp.dat};

            %% gives the temperature of each point where the reference point
            %% is the lower left corner of the cell (also a 4x4 matrix)
            %% (I already asked the package author if this is intended or a bug)
            %\addplot3 [surf,shader=flat corner] file {temp.dat};

            % this should give the desired output (5x5 matrix)

            \addplot [matrix plot*,point meta={abs(\thisrowno{2})}] table [x index=0, y index=1, meta index=2] {figures/data/ErrMatDir_nohier.dat};

            %\addplot [matrix plot*,point meta={abs(\thisrowno{2})}] table [x index=0, y index=1, meta index=2] {figures/data/ErrMatDir6q.dat};
        \end{axis}
        \useasboundingbox (-1,-1.3) rectangle (1,1);
    \end{tikzpicture}
    \end{subfigure}
    \begin{subfigure}[t]{0.3\textwidth}
        \centering
        \caption{\hspace{4cm}\vspace{-0.5cm}}
        \tikzsetnextfilename{ErrMatDir_n6}
\begin{tikzpicture}
\useasboundingbox (-1.5,-1) rectangle (5,4);
        \begin{axis}[
            view={0}{90},   % not needed for `matrix plot*' variant
           ymin=-0.7, ymax=63.4,
            xmin=-0.7, xmax=63.4,
            xtick pos=upper,
            xticklabel pos=upper,
            xlabel={$j$}, 
            width=3cm,
            height=3cm,
            ylabel={$i$}, 
            grid=none,
            y dir=reverse,
            colorbar,
            ytick={0,31,63},
            xtick={0,31,63},
            xticklabels={1,32,64},
            yticklabels={1,32,64},
            xticklabel style = {font=\scriptsize},
            yticklabel style = {font=\scriptsize,xshift=3ex},
            colorbar horizontal,
            colorbar style={
                %title=$^\circ \mathrm{C}$,
                at={(0,-0.1)},
                anchor=north west,
                font =\scriptsize,
                xtick={0.0,0.0005,0.001},                
                % yticklabel style={
                %      /pgf/number format/.cd,
                %      fixed,
                %      precision=2,
                %      fixed zerofill,
                %      },
                scaled ticks=false,
                height=0.1cm,
            },
            %
            % added these key-values
            enlargelimits=false,
            axis on top,
            point meta min=0.00,
            point meta max=0.001,
            %
            % uncomment me to show, that there are really no cells plotted
            % when using the second variant (`surf, shader=flat corner')
            %xmax=5,
            %ymax=5,
        ]
            %% gives the mean temperature of the neighbouring cells (4x4 matrix)
            %\addplot3 [surf] file {temp.dat};

            %% gives the temperature of each point where the reference point
            %% is the lower left corner of the cell (also a 4x4 matrix)
            %% (I already asked the package author if this is intended or a bug)
            %\addplot3 [surf,shader=flat corner] file {temp.dat};

            % this should give the desired output (5x5 matrix)

            \addplot [matrix plot*,point meta={abs(\thisrowno{2})}] table [x index=0, y index=1, meta index=2] {figures/data/ErrMatDir.dat};

            %\addplot [matrix plot*,point meta={abs(\thisrowno{2})}] table [x index=0, y index=1, meta index=2] {figures/data/ErrMatDir6q.dat};
        \end{axis}
        \useasboundingbox (-1,-1.3) rectangle (1,1);
    \end{tikzpicture}
        \label{fig:LaplaceResultsError}
    \end{subfigure}
    \begin{subfigure}[t]{0.3\textwidth}
        \centering
        \caption{\hspace{4cm}\vspace{-0.5cm}}
        \tikzsetnextfilename{ErrMatDir_reg_n6}
\begin{tikzpicture}
\useasboundingbox (-1.68,-1) rectangle (5,4);
        \begin{axis}[
            view={0}{90},   % not needed for `matrix plot*' variant
           ymin=-0.7, ymax=63.4,
            xmin=-0.7, xmax=63.4,
            xtick pos=upper,
            xticklabel pos=upper,
            xlabel={$j$}, 
            width=3cm,
            height=3cm,
            ylabel={$i$}, 
            grid=none,
            y dir=reverse,
            colorbar,
            xticklabel style = {font=\scriptsize},
            yticklabel style = {font=\scriptsize,xshift=3ex},
            colorbar horizontal,
            ytick={0,31,63},
            xtick={0,31,63},
            xticklabels={1,32,64},
            yticklabels={1,32,64},
            colorbar style={
                %title=$^\circ \mathrm{C}$,
                at={(0,-0.1)},
                anchor=north west,
                font =\scriptsize,
                xtick={0.0,0.0005,0.001},                
                %      /pgf/number format/.cd,
                %      fixed,(a) Refernce
                %      precision=2,
                %      fixed zerofill,
                %      },
                scaled ticks=false,
                height=0.1cm,
            },
            %
            % added these key-values
            enlargelimits=false,
            axis on top,
            point meta min=0.0,
            point meta max=0.001,
            %
            % uncomment me to show, that there are really no cells plotted
            % when using the second variant (`surf, shader=flat corner')
            %xmax=5,
            %ymax=5,
        ]
            %% gives the mean temperature of the neighbouring cells (4x4 matrix)
            %\addplot3 [surf] file {temp.dat};

            %% gives the temperature of each point where the reference point
            %% is the lower left corner of the cell (also a 4x4 matrix)
            %% (I already asked the package author if this is intended or a bug)
            %\addplot3 [surf,shader=flat corner] file {temp.dat};

            % this should give the desired output (5x5 matrix)
            %\addplot [matrix plot*,point meta=explicit] file {figures/data/ErrMatDir_reg.dat};
            \addplot [matrix plot*,point meta={abs(\thisrowno{2})}] table [x index=0, y index=1, meta index=2] {figures/data/ErrMatDir_reg.dat};

        \end{axis}
        \useasboundingbox (-1,-1.3) rectangle (1,1);
    \end{tikzpicture}
        \label{fig:LaplaceResultsErrorReg}
    \end{subfigure}
    \vspace{-0.4cm}
    \caption{Application of the learning protocol to the discretized second derivative operator. 
    The ground-truth operator $A$ is shown in (\subref{fig:LaplaceResultsReference}), and the reconstructed operator obtained from the learned block encoding $G$ of experiment C is shown in (\subref{fig:LaplaceResultsLearned}).
    The reconstruction recovers the dominant diagonal structure of the target block $\operatorname{Tr}_{\mathrm{anc}}[P G]$ in the upper-left corner, while also inducing population in the remaining subblocks.
    Panel (\subref{fig:LaplaceResultsCost}) shows the cost-function evolution for the different experiments. 
    Panels (\subref{fig:LaplaceResultsErrorNoHier})--(\subref{fig:LaplaceResultsErrorReg}) show the corresponding absolute residuals $|\mathcal{E}(A,c \,
\text{Tr}_{\text{anc}}[P G])_{\mathrm{A}}|$, $|\mathcal{E}(A,c \,
\text{Tr}_{\text{anc}}[P G])_{\mathrm{B}}|$, and $|\mathcal{E}(A,c \,
\text{Tr}_{\text{anc}}[P G])_{\mathrm{C}}|$, respectively. 
    The results demonstrate that hierarchical learning with regularization suppresses spurious noise and enhances the stability of the recovered operator.}
    \label{fig:LaplaceResults}
\end{figure}

The upper row of Figure~\ref{fig:LaplaceResults} shows the target matrix \(A\) and the learned matrix \(G\) resulting from experiment C (cf.~Figure~\ref{fig:LaplaceResultsLearned}), where the entries $i,j = 1\dots64$ correspond to the encoded target operator $A$. 
The evolution of the objective function \(\mathcal{J}\) for all three experiments is shown in Figure~\ref{fig:LaplaceResultsCost}, where experiment A terminates after approximately $2\times10^{3}$ iterations, whereas experiments B and C terminate after approximately $2\times10^{4}$ iterations.

Experiment A, which employs only global optimization, achieves a reduction of less than two orders of magnitude in the objective function.
The relative error converges to \( \varepsilon(A,c\mathrm{Tr}_{\mathrm{anc}}[PG])_{\mathrm{A}} \approx 0.2790\), while the resulting success probability \( p_A \approx 0.4146 \) remains only slightly above the operator’s intrinsic probability of \(p(A/\|A\|_2) \approx 0.3734\), cf. Equation~\eqref{eq:succ_prob_avg}.
The limited quality of the approximation is further illustrated in Figure~\ref{fig:LaplaceResultsErrorNoHier}, where the absolute residual satisfies \(|\mathcal{E}(A,c \,
\text{Tr}_{\text{anc}}[P G])_{\mathrm{A}}| \sim {10^{-1}} \), cf.~Equation~\eqref{eq:error_rel}.
In contrast, the advantages of the hierarchical optimization strategy become clearly visible in experiments B~and~C, as shown in Figure~\ref{fig:LaplaceResultsCost}. 
The discontinuous jumps in the objective function correspond to increases in the hierarchy level following the termination of Algorithm~\ref{alg:strat} --  line~\ref{alg:strat:op2}. 
Both experiments converge to a stationary point, and the hierarchical optimization strategy allows a reduction of the objective function by approximately five orders of magnitude, corresponding to an improvement of approximately four orders of magnitude relative to experiment A.
The converged relative errors for experiments B~and~C are \( \varepsilon(A,c\mathrm{Tr}_{\mathrm{anc}}[PG])_{\mathrm{B}} \approx  1.72\!\times\!10^{-3} \), and \( \varepsilon(A,c\mathrm{Tr}_{\mathrm{anc}}[PG])_{\mathrm{C}} \approx {7.17\!\times\!10^{-4}} \), respectively. 
Their associated absolute residual is shown in Figure~\ref{fig:LaplaceResultsError} and~\ref{fig:LaplaceResultsErrorReg}.
Experiment C not only achieves lower overall error levels but also demonstrates substantially suppressed outliers. 
This gain in accuracy is achieved without a loss in success probability.
Experiment C achieves a success probability of \(p_C \approx 0.3628\), compared with \(p_B \approx 0.3597\) in experiment B.
Hence, with the stabilizing term active, including the regularization term does not compromise the success
probability.

%%%%%%%%%%%%%%%%%%%%%%%%%%%%%%%%%%%%%%%%%%%%%%%%%%%%%%%%%%%%%%%%%%%%
%%  INVISCID FLOW 
%%%%%%%%%%%%%%%%%%%%%%%%%%%%%%%%%%%%%%%%%%%%%%%%%%%%%%%%%%%%%%%%%%%%
\subsubsection{Potential Flow Analysis}
\label{sec:panel}
\begin{figure}[htbp]
    \centering
    \includegraphics[width=0.7\textwidth]{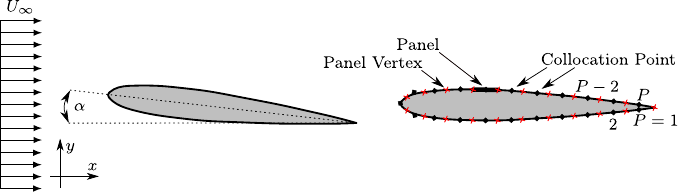}
    \caption{Illustration of the flow around a NACA 0012 airfoil subjected to an approach flow $U_{\infty}$ at AoA $\alpha$ (left) and geometry discretization by $P$ panels with collocation points in the center of each panel (right).}
    \label{fig:panel}
\end{figure}

The final application deals with the steady flow around a two-dimensional NACA 0012 airfoil. 
The airfoil is subject to a uniform free-stream velocity of $U_\infty$ and set at an angle of attack (AoA) $\alpha = 5^\circ$ as depicted in Figure~\ref{fig:panel}.
The underlying physical model is based on an inviscid potential flow, and the numerical algorithm employs a Hess-Smith panel method~\cite{Hess1967}.

The physical model assumes irrotational and solenoidal velocity fields and is typically associated with the limiting case of infinitely large Reynolds numbers, i.e., $Re =U_\infty \bar{c}/\nu \to \infty$, with $\bar{c}$ being the chord length of the airfoil, and $\nu$ the fluid's kinematic viscosity.
Under these assumptions, the velocity vector can be expressed as $\vec{u}=\nabla \phi$, where $\phi$ is a scalar potential field satisfying the Laplace equation $\Delta\phi=0$ as a direct consequence of the conservation of mass.
The linearity of the Laplace equation permits the construction of flow configurations through the superposition of elementary solutions to $\Delta\phi=0$, such as uniform flow as well as sources and free vortices of unknown strength~\cite{Hess1967}. 
The panel method represents the flow around arbitrary geometries as superpositions of such discrete elementary solutions. 
In this study, the potential $\phi$ is decomposed into contributions from the approaching uniform flow $U_\infty$, sources, and a single vortex.
The unknowns to be determined are the strengths of the sources and the vortex. 
These are determined from the kinematic flow condition~\cite{Hess1967} and the Kutta~\cite{Batchelor2000} condition, the former ensures that no flow passes through the body, while the latter guarantees that the flow leaves the trailing edge tangentially. 
For this purpose, the profile geometry is discretized into $P$ panels, with a collocation point defined at the center of each panel where the kinematic flow condition must be satisfied, see Figure~\ref{fig:panel}.
To satisfy the Kutta condition, the magnitudes of the tangential velocities at the upper and lower trailing-edge panels must match.
This leads to a system of linear equations for $P+1$ unknowns ($P$ sources $q_j$ plus one vortex $\gamma$) for a geometry discretized by $P$ panels, viz.
\begin{equation}
        \sum_{j=1}^{P} A_{ij} q_j + A_{i(P+1)} \gamma = b_i  \, :\, \forall i\in[1,P+1]\, .
\end{equation}
Since the elementary solutions essentially correspond to global ansatz functions, the matrix $A$ is fully populated -- in contrast to matrices resulting from typical local ansatz functions. 

\smallskip
To improve our understanding of the complexity of the matrix $A$, we study the singular value spectrum of a suitable partition of the matrix shown. 
Given $A\in\mathbb{C}^{2^{n_\text{s}}\times 2^{n_\text{s}}}$ where $2^{n_\text{s}} = P+1$, we bipartition the ${n_\text{s}}$-qubit Hilbert space into two subsystems of dimension $d=2^{n_\text{s}/2}$. 
The matrix is first reshaped into a rank-four tensor $\mathcal{A}_{i_L,i_R,j_L,j_R} \in \mathbb{C}^{d\times d\times d\times d}$, where $i_L,i_R$ label the row indices and $j_L,j_R$ label the column indices of the two subsystems.
We then realign the tensor by fusing the index pairs $(i_L,j_L)$ and $(i_R,j_R)$, yielding the matrix $A^\sharp_{(i_L,j_L),(i_R,j_R)} = \mathcal{A}_{i_L,i_R,j_L,j_R}$.
The singular-value spectrum $\sigma_i$ studied here is the spectrum of $A^\sharp$, visualized in Figure~\ref{fig:PanelReference:Corr}. 
The method employed corresponds to the Schmidt decomposition or the realignment of $A$ with respect to the bipartition $L|R$. 
For all investigated discretizations $P=2^{n_\text{s}} -1$ with $n_\text{s} \in \{6, 8, 10, 12\}$, a significant number of sufficiently large singular values appear. 
This highlights the complex structure of $A$ and makes its exact encoding into a quantum circuit a challenge.
\begin{figure}[t]
    \centering
    \begin{subfigure}[t]{0.32\textwidth}
        \centering
        \caption{\hspace{3.8cm}\vspace{-0.5cm}}
        \tikzsetnextfilename{Correlation}
\begin{tikzpicture}
\useasboundingbox (-1.48,-1) rectangle (4.2,4.5);
\begin{axis}[
title={\tiny{  }},
log basis y={10},
log basis x={10},
legend style={fill opacity=0.8, draw opacity=1, text opacity=1, at={(0.59,0.5)}},
tick pos=left,
xmajorgrids,
width=3.5cm,
height=3.5cm,
xmin=0, %xmax=485,	
xlabel={$i$}, %{$i$ [\  ]},
ylabel={$\sigma_i$},% [\  ]},
yticklabel style = {font=\scriptsize,xshift=0ex},
xticklabel style = {font=\scriptsize},
ymajorgrids,
% y tick label style={
% text width=2.5em
% },
xmin = 1,
trim axis left,
baseline,
trim axis right,
scale only axis,
%ymin=7.0e-2, 
%ymax=1,
ytick={1e-12,1e-8,1e-4,1e-0},
ymode=log,
xmode=log
]
\addplot  [mark={x},semithick, mark options={mark size=2pt,solid, line width=0.5pt},] table [x expr=\coordindex+1, y index=0] {figures/data/mat6.dat};
\addlegendentry{\scriptsize{$n_\text{s}=6$}}

\addplot  [mark={o},semithick,dashed,mark options={mark size=2pt,solid, line width=0.5pt},] table [x expr=\coordindex+1, y index=0] {figures/data/mat8.dat};
\addlegendentry{\scriptsize{$n_\text{s}=8$}}

\addplot  [mark={star},semithick,dotted,mark options={mark size=2pt,solid, line width=0.5pt},] table [x expr=\coordindex+1, y index=0] {figures/data/mat10.dat};
\addlegendentry{\scriptsize{$n_\text{s}=10$}}

\addplot  [mark={triangle},semithick,dashdotted,mark options={mark size=2pt,solid, line width=0.5pt},] table [x expr=\coordindex+1, y index=0] {figures/data/mat12.dat};
\addlegendentry{\scriptsize{$n_\text{s}=12$}}

\end{axis}
\end{tikzpicture}
        \label{fig:PanelReference:Corr}
    \end{subfigure}
    \begin{subfigure}[t]{0.32\textwidth}
        \centering
        \caption{\hspace{4cm}\vspace{-0.5cm}}
        \tikzsetnextfilename{RefMatPanel}
\begin{tikzpicture}
\useasboundingbox (-1,-1) rectangle (4.2,4.5);
        \begin{axis}[
            %view={0}{90},   % not needed for `matrix plot*' variant
            xlabel={$j$}, 
            width=3.5cm,
            height=3.5cm,
            ymin=-0.7, ymax=31.4,
            xmin=-0.7, xmax=31.4,
            xtick pos=upper,
            xtick style={draw=none},     
            ytick style={draw=none},      
            xticklabel pos=upper,
            ylabel={$i$},
            trim axis left,
            baseline,
            trim axis right,
            scale only axis,
            ytick={0,15,31},
            xtick={0,15,31},
            xticklabels={1,16,32},
            yticklabels={1,16,32},
            grid=none,
            y dir=reverse,
            colorbar,
            xticklabel style = {font=\scriptsize},
            yticklabel style = {font=\scriptsize,xshift=3ex},
            colorbar horizontal,
            colorbar style={
                %title=$^\circ \mathrm{C}$,
                at={(0,-0.1)},
                anchor=north west,
                font =\scriptsize,
                % yticklabel style={
                %     /pgf/number format/.cd,
                %     fixed,
                %     precision=1,
                %     fixed zerofill,
                %     },
                    height=0.1cm,
            },
            %
            % added these key-values
            enlargelimits=false,
            axis on top,
            point meta min=-1.5,
            point meta max=1.5,
            %
            % uncomment me to show, that there are really no cells plotted
            % when using the second variant (`surf, shader=flat corner')
            %xmax=5,
            %ymax=5,
        ]
            %% gives the mean temperature of the neighbouring cells (4x4 matrix)
            %\addplot3 [surf] file {temp.dat};

            %% gives the temperature of each point where the reference point
            %% is the lower left corner of the cell (also a 4x4 matrix)
            %% (I already asked the package author if this is intended or a bug)
            %\addplot3 [surf,shader=flat corner] file {temp.dat};

            % this should give the desired output (5x5 matrix)
            \addplot [matrix plot*,point meta=explicit] file {figures/data/RefMatPanel.dat};
%            \useasboundingbox (-1,-1.3) rectangle (1,1);
        \end{axis}
\end{tikzpicture}
        \label{fig:PanelReference:Mat}
    \end{subfigure}
    \begin{subfigure}[t]{0.32\textwidth}
    \centering
        \caption{\hspace{4cm}\vspace{-0.5cm}}
        \tikzsetnextfilename{LearMatPanel}
\begin{tikzpicture}
\useasboundingbox (-1,-1) rectangle (4.2,4.5);
        \begin{axis}[
            view={0}{90},   % not needed for `matrix plot*' variant
            xlabel={$j$},      
            xtick pos=upper,
            width=3.5cm,
            height=3.5cm,
            xticklabel pos=upper,
            xtick style={draw=none},     
            ytick style={draw=none},      
            xticklabel pos=upper,
            ylabel={$i$},
            trim axis left,
            baseline,
            trim axis right,
            scale only axis,
            ytick={0,15,31,63},
            xtick={0,15,31,63},
            xticklabels={1,16,32,64},
            yticklabels={1,16,32,64},
            grid=none,
            y dir=reverse,
            colorbar,
            xticklabel style = {font=\scriptsize},
            yticklabel style = {font=\scriptsize,xshift=3ex},
            colorbar horizontal,
            colorbar style={
                %title=$^\circ \mathrm{C}$,
                at={(0,-0.1)},
                anchor=north west,
                font =\scriptsize,
                % yticklabel style={
                %     %/pgf/number format/.cd,
                %     %fixed,
                %     %precision=1,
                %     fixed zerofill,
                %     },
                    height=0.1cm,
            },
            %
            % added these key-values
            enlargelimits=false,
            axis on top,
            point meta min=-1.50,
            point meta max=1.50,
            %
            % uncomment me to show, that there are really no cells plotted
            % when using the second variant (`surf, shader=flat corner')
            %xmax=5,
            %ymax=5,
        ]
            %% gives the mean temperature of the neighbouring cells (4x4 matrix)
            %\addplot3 [surf] file {temp.dat};

            %% gives the temperature of each point where the reference point
            %% is the lower left corner of the cell (also a 4x4 matrix)
            %% (I already asked the package author if this is intended or a bug)
            %\addplot3 [surf,shader=flat corner] file {temp.dat};

            % this should give the desired output (5x5 matrix)
            \addplot [matrix plot*,point meta=explicit] file {figures/data/LearMatPanel.dat};
        \end{axis}
\useasboundingbox (-0.2,-1) rectangle (1,1);
    \end{tikzpicture}
        \label{fig:PanelResultsLearned}
    \end{subfigure}
    \caption{Sparsity and correlation measures of the potential flow operator $A\in\mathbb{R}^{P+1\times 
    P+1}$.  
    Subfigure (\subref{fig:PanelReference:Corr}) depicts the evolution of singular values $\sigma_i$ of the central partition matrix $A^\sharp$ for multiple discretizations  $P=2^{n_\text{s}} -1$. 
    Subfigures (\subref{fig:PanelReference:Mat}, \subref{fig:PanelResultsLearned}) show the target matrix $A$ for a system size $P+1=32$ as well as the learned operator $G$ in its block-encoded version (regularization disabled), with the approximation to $A$ located in the upper-left corner.}
    \label{fig:PanelReference}
\end{figure}

Figure~\ref{fig:PanelResultsLearned} displays the learned matrix $G$ for a discretization with $P=31$ panels, where $16$ panels are located on the upper side (suction side) and $15$ panels on the lower side (pressure side) of the airfoil.
The dense structure of the target matrix $A$ is shown in Figure~\ref{fig:PanelReference:Mat} and requires $n_\text{s}=5$ and $n_\text{a}=1$ qubits for encoding.
 
The optimization is performed using the L-BFGS optimizer setting stepsize $\eta=0.02$, maximum number of iterations $\texttt{maxit}\sim10^3$ for each of the hierarchical learning sequences in Algorithm~\ref{alg:strat} and $\texttt{maxit}\sim 4\!\times\!10^3$ for the final training executed in Algorithm~\ref{alg:strat} -- line~\ref{alg:strat:op3}. 
Furthermore, the convergence criterion is defined on the gate-wise derivative of the cost function, cf. Equation ~\eqref{eq:saddle}, as $\| \partial\mathcal{J}/\partial G\| \leq \texttt{tol} = 10^{-15}$. 
Two experiments were conducted for demonstration purposes: one with regularization disabled ($\rho=0$, experiment A) and one with regularization enabled ($\rho=10^{-6}$, experiment B).
In both cases, the stabilizer for the success probability was set to $\mu=0.0$.

The quantum circuit is constructed out of $l=4$ layers, where gate primitives are set to nearest-neighbor $r=r^j=4$-qubit gates. This corresponds to the gate sequence for LNN-like chip topologies, which has been used in the previous experiment and is visualized in Figure~\ref{fig:LaplaceLNN}.
Additional hardware topology constraints along the learning are not imposed, such that
\begin{equation}
        G = \prod^{4}_{j=1}\prod^{n-3}_{i=1}G^j_i, \quad \mathcal{G}^j_i = (i,i+1,i+2,i+3)\,.
\end{equation}

The learned operator $G$ in its block-encoded version is shown in Figure~\ref{fig:PanelResultsLearned}, in which the entries $i,j = 0\dots31$ correspond to an approximation to the target matrix $A$ with regularization disabled ($\rho =0$).
The convergence of the normalized cost function and the relative errors obtained with and without regularization are displayed in Figure~\ref{fig:PanelResults}.
The convergence history of the objective functional displays small fluctuations, which are due to the Wolfe-Conditions employed within the L-BFGS optimizer, and a greater descent in experiment B.
Regardless of whether regularization is used, the cost function $\mathcal{J}$ decreases by more than $15$ orders of magnitude.
The use of amplitude encoding implies long-range qubit interactions, as the correlations of the target operator are not aligned with those of the qubit register. 
Despite this mismatch, the presented method achieves high accuracy in the learned operator with the relative error of $\varepsilon(A,c\text{Tr}_{\text{anc}}[P G])_{\mathrm{A}} \approx 1.78\!\times\!10^{-9}$ (regularization disabled) and  $\varepsilon(A,c\text{Tr}_{\text{anc}}[P G])_{\mathrm{B}} \approx 4.10 \!\times\! 10^{-10}$ (regularization with $\rho=10^{-6}$), cf.~Figures~\ref{fig:PanelResultsError} and \ref{fig:PanelResultsErrorReg}. 
These figures show that the inclusion of the regularization term provides an improvement, w.r.t. the maximum relative error.
Furthermore, a substantial reduction of the outliers, which reduces the relative error $\varepsilon(A,c\text{Tr}_{\text{anc}}[P G])$, is observed in the results.
For practical calculations, the averaged success probabilities outlined in Equation~\eqref{eq:succ_prob_avg} are relevant. 
These also improve as a result of regularization. 
Using a single ancilla qubit ($n_a=1$), they stand at $p\approx 0.02766$ without regularization and $p \approx  0.04824$ with regularization. 
Generally, the experiments confirm the learning of dense and strongly correlated operators, and highlight potential to encode other structured dense operators arising in classical scientific computing. 
\begin{figure}[t]
\centering
      \begin{subfigure}[t]{0.34\textwidth}
            \centering
            \caption{\hspace{3.5cm}\vspace{-0.5cm}}
            \input{figures/Panel_Cost}
        \label{fig:PanelResultsCost}
    \end{subfigure}
      \begin{subfigure}[t]{0.32\textwidth}
        \centering
        \caption{\hspace{4cm}\vspace{-0.5cm}}
        \tikzsetnextfilename{ErrMatPanel}
\begin{tikzpicture}
\useasboundingbox (-1,-1) rectangle (4.2,4.5);
        \begin{axis}[
            view={0}{90},   % not needed for `matrix plot*' variant
            xlabel={$j$},
            ymin=-0.7, ymax=31.4,
            xmin=-0.7, xmax=31.4,
            xtick pos=upper,
            xticklabel pos=upper,
            width=3.5cm,
            height=3.5cm,
            xtick style={draw=none},     
            ytick style={draw=none},      
            xticklabel pos=upper,
            ylabel={$i$},
            trim axis left,
            baseline,
            trim axis right,
            scale only axis,            
            grid=none,
            y dir=reverse,
            colorbar,
            ytick={0,15,31},
            xtick={0,15,31},
            xticklabels={1,16,32},
            yticklabels={1,16,32},
            xticklabel style = {font=\scriptsize},
            yticklabel style = {font=\scriptsize,xshift=3ex},
            colorbar horizontal,
            colorbar style={
                %title=$^\circ \mathrm{C}$,
                at={(0,-0.1)},
                anchor=north west,
                font =\scriptsize,
                xtick={0.000000000,0.00000000015,0.0000000003},
                %xtick={-0.0000000004,0,0.0000000004},
                % yticklabel style={
                %      /pgf/number format/.cd,
                %      fixed,
                %      precision=2,
                %      fixed zerofill,
                %      },
                scaled ticks=false,
                height=0.1cm,
            },
            %
            % added these key-values
            enlargelimits=false,
            axis on top,
            %point meta min=-0.0000000004,
            point meta min=-0.0000000,
            point meta max=0.0000000003,
            %
            % uncomment me to show, that there are really no cells plotted
            % when using the second variant (`surf, shader=flat corner')
            %xmax=5,
            %ymax=5,
        ]
            %% gives the mean temperature of the neighbouring cells (4x4 matrix)
            %\addplot3 [surf] file {temp.dat};

            %% gives the temperature of each point where the reference point
            %% is the lower left corner of the cell (also a 4x4 matrix)
            %% (I already asked the package author if this is intended or a bug)
            %\addplot3 [surf,shader=flat corner] file {temp.dat};

            % this should give the desired output (5x5 matrix)
            \addplot [matrix plot*,point meta={abs(\thisrowno{2})}] table [x index=0, y index=1, meta index=2] {figures/data/ErrMatPanel.dat};
        \end{axis}
       % \useasboundingbox (-1,-1.3) rectangle (1,1);
    \end{tikzpicture}
    \label{fig:PanelResultsError}
    \end{subfigure}
    \begin{subfigure}[t]{0.32\textwidth}
        \centering
        \caption{\hspace{4cm}\vspace{-0.5cm}}
        \tikzsetnextfilename{ErrMatPanelRegularizer}
\begin{tikzpicture}
\useasboundingbox (-1,-1) rectangle (4.2,4.5);
        \begin{axis}[
            view={0}{90},   % not needed for `matrix plot*' variant
            xlabel={$j$},
            ymin=-0.7, ymax=31.4,
            xmin=-0.7, xmax=31.4,
            xtick pos=upper,
            xticklabel pos=upper,
            width=3.5cm,
            height=3.5cm,
            xtick style={draw=none},     
            ytick style={draw=none},      
            xticklabel pos=upper,
            ylabel={$i$},
            trim axis left,
            baseline,
            trim axis right,
            scale only axis,            
            grid=none,
            y dir=reverse,
            colorbar,
            ytick={0,15,31},
            xtick={0,15,31},
            xticklabels={1,16,32},
            yticklabels={1,16,32},
            xticklabel style = {font=\scriptsize},
            yticklabel style = {font=\scriptsize,xshift=3ex},
            colorbar horizontal,
            colorbar style={
                %title=$^\circ \mathrm{C}$,
                at={(0,-0.1)},
                anchor=north west,
                font =\scriptsize,
                xtick={0.000000000,0.00000000015,0.0000000003},
               % xtick={-0.0000000004,0,0.0000000004},    
                % yticklabel style={
                %      /pgf/number format/.cd,
                %      fixed,
                %      precision=2,
                %      fixed zerofill,
                %      },
                scaled ticks=false,
                height=0.1cm,
            },
            %
            % added these key-values
            enlargelimits=false,
            axis on top,
%            point meta min=-0.0000000004,
            point meta min=-0.000000,
            point meta max=0.0000000003,
            %
            % uncomment me to show, that there tare really no cells plotted
            % when using the second variant (`surf, shader=flat corner')
            %xmax=5,
            %ymax=5,
        ]
            %% gives the mean temperature of the neighbouring cells (4x4 matrix)
            %\addplot3 [surf] file {temp.dat};

            %% gives the temperature of each point where the reference point
            %% is the lower left corner of the cell (also a 4x4 matrix)
            %% (I already asked the package author if this is intended or a bug)
            %\addplot3 [surf,shader=flat corner] file {temp.dat};

            % this should give the desired output (5x5 matrix)
            %\addplot [matrix plot*,point meta=explicit] file {figures/data/ErrMatPanel_regularizer.dat};
             \addplot [matrix plot*,point meta={abs(\thisrowno{2})}] table [x index=0, y index=1, meta index=2] {figures/data/ErrMatPanel_regularizer.dat};
        \end{axis}
        %\useasboundingbox (-1,-1.3) rectangle (1,1);
    \end{tikzpicture}
    \label{fig:PanelResultsErrorReg}
    \end{subfigure}
    \caption{Results for the learning of an inviscid airfoil flow operator showing the evolution of the cost functional  (\subref{fig:PanelResultsCost}) -- without $\rho=0$ (red) and with $\rho=10^{-6}$ (black) regularizer -- and the absolute residual $|\mathcal{E}(A,c \,\text{Tr}_{\text{anc}}[P G])|$ to the target matrix $A$ without (\subref{fig:PanelResultsError}) and with regularization (\subref{fig:PanelResultsErrorReg}).
    }
    \label{fig:PanelResults}
\end{figure}

%%%%%%%%%%%%%%%%%%%%%%%%%%%%%%%%%%%%%%%%%%%%%%%%%%%%%%%%%%%%%%%%%%%%
%%  Conclusions
%%%%%%%%%%%%%%%%%%%%%%%%%%%%%%%%%%%%%%%%%%%%%%%%%%%%%%%%%%%%%%%%%%%%
\section{Conclusion}
\label{sec:conclusions}
In this work, we introduced a hardware-agnostic variational operator-learning framework for compiling discrete operators into compact quantum circuits.
The method is designed to treat both unitary and non-unitary operators, and to incorporate hardware constraints directly at the circuit-construction level, including restricted qubit connectivity, long-range gates, and multi-qubit interaction patterns. 
To this end, the algorithm maps a target operator to a trained circuit, minimizing the normalized Frobenius error between the two operators during training.
The protocol integrates a backpropagation-based optimization strategy combined with manifold retraction capable of learning unitary operators directly, while non-unitary operators are learned through a cost-efficient block encoding.
Unlike methods whose ancilla number scales with auxiliary dimension or LCU terms, the proposed block encoding uses only one global ancilla for non-unitary operators. 
Finally, the hierarchically executed learning process can incorporate a regularizer for smoothing residual errors in order to suppress localized outliers and improve the learned amplitude-encoded operators.

\smallskip
We quantitatively assess the generality of our programming protocol across a range of application domains, including quantum physics (many-body dynamics), quantum chemistry (butadiene molecule), and engineering (finite-difference approximations, potential flow analysis). 
These examples demonstrate that (a) the method yields compact circuit representations with higher accuracy compared to published reference techniques, (b) operator learning can serve as a practical programming layer between classical discretized models and quantum hardware, and (c) the protocol offers powerful capabilities for circuit compression, thereby effectively adapting resource consumption in quantum computations to specified constraints.

In conjunction with quantum applications, the proposed framework is evaluated using unitary time-evolution operators, where it outperforms standard Suzuki-Trotter decompositions and exhibits lower approximation errors in all experiments. 
The largest performance gains are observed for Hamiltonians exhibiting pronounced structure and locality, where the ansatz can exploit correlations that are not efficiently captured by fixed-order Trotter schemes. 
At the same time, the results indicate that circuit quality depends heavily on the initialization strategy, highlighting initialization as a crucial component.

For all engineering applications examined here, accuracy close to machine precision is achieved up to the $5$-qubit instance, while requiring fewer ancilla qubits than TN-based constructions. 
The framework successfully learned both sparse, structured and non-sparse (dense), unstructured operators. 
This suggests that the approach is not restricted to sparse Hamiltonian operators and motivates its application to more general operators, for instance, those associated with the unstructured computational grids encountered in classical scientific computing.
A principal limitation observed, particularly for the approximation of the Laplacian, is the lack of a scalable initialization strategy. Consequently, this remains an important aspect for future work. 

\smallskip
The algorithm could be improved and extended in various ways. First, scalability beyond the current implementation in the full Hilbert space requires an extension based on TNs, see, e.g., Reference~\cite{Gibbs2025}.
Such a formulation would reduce the classical memory and contraction cost, at the price of restricting the method to operators that admit efficient TN compression.
Second, the circuit ansatz could be made adaptive by adding gates according to gradient-based criteria, in the spirit of ADAPT-VQAs~\cite{Grimsley2019}.
Related adaptive strategies have recently shown strong performance in quantum state learning~\cite{Jaderberg2026}, and could provide a systematic way to allocate circuit resources only where they are most effective.
Third, versatility can be further enhanced by investigating retraction alternatives such as QR-decomposition-based approaches (Householder/Givens projections or Newton–Schulz iteration), and introducing an adaptive learning rate as well as derivative-dependent regularization terms. 
Finally, while qubit connectivity and device layout already enter the ansatz through the gate supports $\mathcal{G}_i^j$, the native gate set and the gate sizes $r^j$ are currently fixed prior to the optimization.
Including these choices directly into the optimization problem, e.g., by adjusting connectivity lists induced by a given qubit assignment, would extend the presented algorithm to incorporate non-trivial logical-to-physical qubit maps, which constitutes an important research direction.
Beyond that, noise-aware optimization, in which the dissipative channels of the respective hardware noise model are incorporated directly into the optimization procedure, is a promising direction for future work.

\smallskip
In conclusion, the four test cases presented here demonstrate the applicability of our method to non-trivial operators arising in quantum computing and computational engineering.
Extending it to a broader class of operator programming problems is a promising direction for future research. 

%%%%%%%%%%%%%%%%%%%%%%%%%%%%%%%%%%%%%%%%%%%%%%%%%%%%%%%%%%%%%%%%%%%%
%%  Acknowledgments
%%%%%%%%%%%%%%%%%%%%%%%%%%%%%%%%%%%%%%%%%%%%%%%%%%%%%%%%%%%%%%%%%%%%
\begin{acknowledgments}
This publication and the current work have received funding from the \href{https://doi.org/10.3030/101080085}{European Union's Horizon Europe research and innovation program (HORIZON-CL4-2021-DIGITAL-EMERGING-02-10) under grant agreement No. 101080085 \textsc{QCFD}}. 
\end{acknowledgments}

%###########################################
%       Data Availability
%###########################################
\section*{Data availability}
\addcontentsline{toc}{section}{Data availability}
The data that support the findings of this study are openly available at the following URL/DOI: \href{https://doi.org/10.25592/uhhfdm.18962}{doi.org/10.25592/uhhfdm.18962}.

%###########################################
%       Appendix
%###########################################
{
\appendix
\renewcommand\thesubsection{\thesection.\arabic{subsection}}
\section{Spectral-to-Frobenius ratio of the residual}
\label{app:specfro}
The relative error $\varepsilon(A,B)$, defined in Equation~\eqref{eq:relativeErrorScalar} and used throughout this work to assess accuracy, is built on the Frobenius norm.
This norm measures the average, and not the worst-case, deviation between the target $A$ and the learned operator $B$.
For example, a residual can have a small Frobenius norm while still acting almost maximally wrongly on one specific input state.
For a quantitative demonstration, we take the Frobenius norm of the residual in Equation~\eqref{eq:error_rel}, where the residual is $\mathcal{E}(A,B) = A - B$ with $B = c\,\text{Tr}_\text{anc}[PG]$. 
To this end, assume $A=\mathbb{I}$, $B=\mathbb{I}-\ket{0}\bra{0}$ of size $2^n\times 2^n$, for which the error is $\varepsilon(A,B)=1/\sqrt{2^n}$ and obviously decreasing with rising system size.
This issue gets set into relation by the norm inequality \cite{Golub2013}, viz.
\begin{equation}
    \frac{\|\mathcal{E}(A,B)\|_\mathrm{fro}}{\sqrt{N}}
    \;\leq\; \|\mathcal{E}(A,B)\|_2
    \;\leq\; \|\mathcal{E}(A,B)\|_\mathrm{fro}\,.
    \label{eq:fro_spec_inequality}
\end{equation}
Herein, the worst-case error is governed by the spectral norm 
\begin{equation}
    \|\mathcal{E}(A,B)\|_2
    = \max_{\||\psi\rangle\|=1}
      \big\|\mathcal{E}(A,B)\,|\psi\rangle\big\|\,,
    \label{eq:spec_worst_case}
\end{equation}
where $\|\cdot\|$ denotes the Euclidean norm of a vector.
Equation~\eqref{eq:spec_worst_case} makes the dependence on the input state explicit, as the spectral norm is the error for the least favorable input state.
The lower bound in Equation~\eqref{eq:fro_spec_inequality} is Frobenius-norm based and corresponds to the average over input states.
Averaging over Haar-random input states gives~\cite{Mele2024}
\begin{equation}
    \mathbb{E}_{\psi}\Big[\,
      \big\|\mathcal{E}(A,B)\,|\psi\rangle\big\|^{2}\,\Big]
    = \frac{\operatorname{Tr}\!\big(
        \mathcal{E}^\dagger(A,B)\,\mathcal{E}(A,B)\big)}{N}
    = \frac{\|\mathcal{E}(A,B)\|_\mathrm{fro}^{2}}{N}\,.
    \label{eq:haar_average}
\end{equation}
Thus, $\|\mathcal{E}(A,B)\|_\mathrm{fro}/\sqrt{N}$ is the root-mean-square error over input states.
Equation~\eqref{eq:fro_spec_inequality} is therefore the statement that the worst case is never below, and at most $\sqrt{N}$ times above, the average measure. 

As a consequence of the discussion above, we employ the ratio of the two measures to relate the average error to the worst-case scenario, viz.
\begin{equation}
    \zeta(A,B)
    = \frac{\|\mathcal{E}(A,B)\|_2}
           {\|\mathcal{E}(A,B)\|_\mathrm{fro}\big/\sqrt{2^{n_\text{s}}}}
    = \frac{\sqrt{2^{n_\text{s}}}\;\|\mathcal{E}(A,B)\|_2}
           {\|\mathcal{E}(A,B)\|_\mathrm{fro}}\,.
    \label{eq:specfro_ratio}
\end{equation}
This ratio $\zeta$ measures how far the worst-case exceeds the average, and hence how uniformly the residual is spread over input states.
For the sake of completeness, $\zeta = 1$ is reached when the singular values of $\mathcal{E}(A,B)$ are uniform, whereas $\zeta = \sqrt{N}$ when $\mathcal{E}(A,B)$ is rank one, which recovers the exemplary given situation above.

For clarity, we therefore report $\zeta$ for the optimal (converged) result of each application presented in Section~\ref{sec:result}, within the next sections.

\subsection{Quantum Applications}
\label{app:specfro:quantum}
Figure~\ref{fig:specfro:ising} shows $\zeta(U(\delta t),*)$ for the three trainable ansätze $G_1$, $G_2$, and $G_3$ as a function of the number of system qubits $n$, with the time step fixed at $\delta t=0.1$. 
These are the same simulations shown in Figure~\ref{fig:HamSimQubitError}. 
Over the system sizes tested, $\zeta$ increases only weakly and appears to scale subexponentially with $n$, in contrast to the exponential scaling of its upper bound.
This suggests that error concentration does not become more pronounced over the tested range.

Figure~\ref{fig:specfro:ppp} shows $\zeta(U(\delta t),G)$ for the butadiene PPP propagator as a function of the time step $\delta t$, with $n=8$ fixed. 
These are the same simulations shown in Figure~\ref{fig:qchemerror}.
Here, $\zeta$ remains approximately constant as $\delta t$ varies, indicating no increase in error
concentration over the time steps tested.
\begin{figure}[t]
\centering
    \vspace{-0.5cm}
    \begin{subfigure}[t]{0.48\textwidth}
       \centering
        \caption{\hspace{6.7cm}\vspace{-0.8cm}}
        \tikzsetnextfilename{SpecFro_Ising}
\begin{tikzpicture}
    \begin{semilogyaxis}[
        xlabel={$n$},% [\  ]},
        ylabel={$\zeta(U(\delta t),*)$ },
        width=6cm,
        height=3cm,
        xtick={4,6,8,10,12},%,14,16},
        scale only axis,
        xmin=3, xmax=13,
        yminorgrids=true,
        %ymin=1e-2, ymax=1e-0,
        legend style={
            at={(0.5,1.03)},
            anchor=south,
            legend columns=4,
            /tikz/every even column/.append style={
                column sep=0.5cm},
        },
        ]
    
    % n qubits two qubit -  1 layer -- Not covered by termanova -- corresponds to trotter 1st 
    \addplot[mark=x] coordinates {
        (4, 1.4099562217094168)
        (6, 1.9905576220460415)
        (8, 2.436924389249903)
        (10, 2.8134600757936186) 
        (12, 3.1452866176895076)
    };
        \addlegendentry{\scriptsize$G_1$}

    %n qubits two qubit -  2 layer -- Not covered by termanova  - corresponds to trotter 2nd
    \addplot[mark=o,mark options={solid}] coordinates {
        (4, 1.3732542487354376 ) 
        (6, 2.0373769493162253)
        (8, 2.326132358340763  ) 
        (10, 2.4377297556106794)
        (12, 2.679934705175298)
    };
           \addlegendentry{\scriptsize$G_2$}

    % n qubits three qubit Termanova R=2
    \addplot[mark=triangle, dashed, mark options={solid}] coordinates {
        (4, 1.0002016129978768) 
        (6, 1.7287653690640612) 
        (8, 2.2272391145635293) 
        (10, 2.6332399360424392) 
        (12, 2.98464003455053)
    };
    \addlegendentry{\scriptsize$G_3$}

    \addplot[red, dotted, domain=2:14, samples=5]{sqrt(2^x)};
    \addlegendentry{\scriptsize$\sqrt{2^n}$}

\end{semilogyaxis}
\end{tikzpicture}
        \label{fig:specfro:ising}
    \end{subfigure}
        \hspace{0.1cm}
    \begin{subfigure}[t]{0.48\textwidth}
        \centering
        \caption{\hspace{6.7cm}\vspace{-0.8cm}}
        \tikzsetnextfilename{SpecFro_PPP}
\begin{tikzpicture}
    \begin{loglogaxis}[
        xlabel={$\delta t$},% [\  ]},
    	ylabel={},
        xmin=0.01, xmax=1,
        ymin=1,
        width=6cm,
        height=3cm,
        legend style={
            at={(0.5,1.03)},
            anchor=south,
            legend columns=2,
            /tikz/every even column/.append style={column sep=0.5cm}
        },
        scale only axis,
         yticklabel style={
            /pgf/number format/fixed,
            /pgf/number format/precision=0,
        },
        yminorgrids=true,
        ytick={1e+0,1e1},
        ]
    
    % n qubits two qubit -  1 layer -- Not covered by termanova -- corresponds to trotter 1st 
    % n =  8 qubits 1 layer 3 qubits + 3 layers two qubit (long range) 
     \addplot[mark=x] coordinates {
          (0.01, 2.0911140865121713) %U2 and U3 opt  0.001981937005365251
          (0.02, 2.0738857632016487) %U2 and U3 opt   0.007804072561877454
          (0.05, 2.1126762036783098) %U2 and U3 opt  0.04434125950755523
          (0.1, 2.221261673263527) %U2 and U3 opt   0.14137912293402505
          (0.2, 2.2210255367017284)  %U2 and U3 opt  0.4055546512408584
          (0.5, 1.337285991496892) %U2 and U3 opt   
          (1.0, 1.602493350136631) %U2 and U3 opt   0.8606481405334323
     };
        \addlegendentry{\scriptsize$G$}

    \addplot[red, dotted, domain=0.01:1, samples=7]{sqrt(2^8)};
    \addlegendentry{\scriptsize$\sqrt{2^8}$}

\end{loglogaxis}
\end{tikzpicture} 
        \label{fig:specfro:ppp}
    \end{subfigure}
    \caption{Spectral-to-Frobenius ratio $\zeta(U(\delta t),*)$ for the quantum applications, cf. Equation~\eqref{eq:specfro_ratio}. Quantum Simulation results are shown in Subfigure (\subref{fig:specfro:ising}), as a function of system size $n$ at fixed $\delta t = 0.1$, for ansätze $G_1$, $G_2$, $G_3$, whereas results for the Quantum Chemistry application, given in Subfigure (\subref{fig:specfro:ppp}), maintain a constant system size $n=8$ demonstrating the error for a changing $\delta t$.}
    \label{app:fig:specfro:quantum}
\end{figure}

\subsection{Engineering Applications}
\label{app:specfro:eng}
Figure~\ref{fig:specfro:laplace} shows the spectral-to-Frobenius ratio $\zeta\!\left(A,c\,\operatorname{Tr}_{\mathrm{anc}}[PG]\right)$ for the discretized second-derivative operator. 
The ratio is plotted as a function of system size $n_{\mathrm{s}}$ for the LNN and 2DL ansätze, using the simulations shown in Figure~\ref{fig:LaplaceQubitError}. 
Although the ratio is considerably smaller than its upper bound, the data are consistent with superpolynomial scaling. 
Oscillations in the numerical results, however, prevent a reliable fit. 
No regularization is used in these simulations ($\rho=0$).

Table~\ref{app:specfro:panel} reports the spectral-to-Frobenius ratio for the panel-method operator without ($\rho=0$) and with ($\rho= 10^{-6}$) regularization, cf. Section~\ref{sec:panel}.
In this example, regularization reduces both the average relative error and $\zeta$.
\begin{figure}[t]
\centering
    \vspace{-0.5cm}
    \begin{subfigure}[t]{0.48\textwidth}
       \centering
        \caption{\hspace{5cm}\vspace{-0.1cm}}
        \tikzsetnextfilename{SpecFro_Laplace}
\begin{tikzpicture}
\begin{semilogyaxis}[
        xlabel={$n_\text{s}$},% [\  ]},
        ylabel={$\zeta(A,c\text{Tr}_\text{anc}[PG])$},
        width=6cm,
        height=3cm,
        scale only axis, 
        xtick={4,5,6,7,8,9},
        legend style={
            at={(0.5,1.03)},
            anchor=south,
            legend columns=3,
            /tikz/every even column/.append style={
                column sep=0.5cm},
        },
        ]

    % LNN  
    \addplot[mark=x] coordinates {
        (4, 1.8485456684079402)
        (5, 3.682068344330189)  
    	(6, 2.668051950001531) 
        (7, 2.9813397993181234) 
        (8, 7.368285728663991)%
        (9, 4.824197260562068)
   	    %(10, 10.0) 
    };
    \addlegendentry{\scriptsize LNN}

      % 2D  
    \addplot[mark=o,orange] coordinates {
        (4, 2.1307321273025943) 
        (5, 2.3844238041207664)  
    	(6, 2.1008208520903144) 
        (7, 4.610657517531131)  
    	(8, 7.2996631154077205)
    	(9, 9.74479766494825)
%   	(10, 1.0)
    };
        \addlegendentry{\scriptsize 2DL}

    \addplot[red, dotted, domain=4:9, samples=5]{sqrt(2^x)};
    \addlegendentry{\scriptsize$\sqrt{2^{n_\text{s}}}$}

%\draw (axis cs: 7.5, 1e-9) node{\usebox{\myboxxx}};
         
\end{semilogyaxis}
\end{tikzpicture} 
        \label{fig:specfro:laplace}
    \end{subfigure}
        \hspace{0.1cm}
    \begin{subfigure}[t]{0.48\textwidth}
        \centering
        \caption{\hspace{5cm}\vspace{-0.1cm}}
        \vspace{0.7cm}
        \begin{tabular}{@{}lcc@{}}
        \toprule
              & $\rho = 0$                 & $\rho = 10^{-6}$                   \\ \midrule
        $\varepsilon\approx$ & $1.78\!\times\! 10^{-9}$ & $4.10 \!\times\! 10^{-10}$ \\
        $\zeta\approx$  & 2.19    & 2.08    \\ \bottomrule
        \end{tabular}
        \label{app:specfro:panel}
    \end{subfigure}
    \vspace{-0.5cm}
    \caption{
    Spectral-to-Frobenius ratio $\zeta(A,c\,\text{Tr}_\text{anc}[PG])$ for the engineering applications, cf. Equation~\eqref{eq:specfro_ratio}.
    Simulation results for the discretized second-derivative operator are shown in Subfigure (\subref{fig:specfro:laplace}), as a function of system size $n_\text{s}$, for LNN and 2DL ansätze, whereas results for the potential flow analysis application with and without regularization are given in Table (\subref{app:specfro:panel}).}
    \label{app:fig:specfro:eng}
\end{figure}
\bigskip

Overall, $\zeta$ provides a simple measure of error concentration. 
In most of the cases tested, minimizing the average error yields worst-case and typical errors that do not differ substantially.
}

{
\appendix
\setcounter{section}{1}
\renewcommand\thesubsection{\thesection.\arabic{subsection}}
\section{Influence of the normalization factor}
\label{app:influence_c}
This appendix illustrates the role of the stabilizing term $\mu c^2$ in Equation~\eqref{eq:control_problem_unconstrained} for the discretized second-derivative operator of Section~\ref{sec:Laplace} with $n_\text{s}=6$.
For this operator, $c_\text{ideal} = \|A\|_2 \approx 3.998$.
If the residual is small, $A \approx c\,\text{Tr}_\text{anc}[PG]$, the average success probability, Equation~\eqref{eq:succ_prob_avg}, behaves as
\begin{equation}
    p(PGP) \approx \frac{\|A\|_\mathrm{fro}^2}{2^{n_\text{s}}\,c^2}
    = p\big(A/\|A\|_2\big)\,\Big(\frac{c_\text{ideal}}{c}\Big)^2 \, ,
    \label{eq:p_vs_c}
\end{equation}
with the operator-intrinsic value $p(A/\|A\|_2) \approx 0.3734$, cf.\ Figure~\ref{fig:LaplaceSucProb}.
Any increase of $c$ beyond $c_\text{ideal}$ therefore reduces the success probability quadratically.

Table~\ref{tab:influence_c} compares four configurations, using the same optimization settings as in Section~\ref{sec:Laplace}:
(A) no hierarchical optimization, (B$_0$) hierarchical optimization without the stabilizing term ($\mu = 0$), (B) hierarchical optimization with $\mu = 10^{-6}$, and (C) hierarchical optimization with $\mu = 10^{-6}$ and regularization ($\rho = 4\!\times\!10^{-8}$).
Configurations A, B, and C correspond to the experiments shown in Figure~\ref{fig:LaplaceResults}.
\begin{table}[h]
\centering
\caption{Normalization factor $c$, relative error $\varepsilon(A, c\,\text{Tr}_\text{anc}[PG])$ and average success probability $p$ for the discretized second-derivative operator with $n_\text{s} = 6$. 
The operator-intrinsic values are $c_\text{ideal} \approx 3.998$ and $p(A/\|A\|_2) \approx 0.3734$.}
\label{tab:influence_c}
\begin{tabular}{l c c c c c}
\toprule
Configuration & $\mu$ & $\rho$ & $c\approx$
  & $\varepsilon\approx$ & $p\approx$ \\
\midrule
A: no hierarchical          & $10^{-6}$ & $0$ & $3.64$ &  $2.79\!\times\!10^{-1}$ & $0.4146$ \\
B$_0$: hierarchical     & $0$       & $0$ & $96.74$ & $2.07\!\times\!10^{-3}$ &$ 0.0006$ \\
B: hierarchical         & $10^{-6}$ & $0$ & $4.07$ & $1.72\!\times\!10^{-3}$ & $0.3597$ \\
C: hierarchical + regularized & $10^{-6}$ & $4\!\times\!10^{-8}$ & $4.06$ & $7.17\!\times\!10^{-4}$ & $0.3628$\\
\bottomrule
\end{tabular}
\end{table}

For the inaccurate result of configuration A, $c$ falls slightly below $c_\text{ideal}$ and, consequently, $p$ slightly exceeds the operator-intrinsic value.
This is possible because $c\,\text{Tr}_\text{anc}[PG]$ is only an approximation to the discretized second-derivative operator, whereas the bound $c \geq c_\text{ideal}$ applies to exact
encodings ($\tau=0$), cf. Section~\ref{sec:math}.
Without the stabilizing term (B$_0$), the optimization converges to a normalization factor $c \gg c_\text{ideal}$ and, in accordance with Equation~\eqref{eq:p_vs_c}, to a correspondingly small success probability.
Since the residual depends only on the product $c\,\text{Tr}_\text{anc}[PG]$, a smaller block combined with a larger $c$ reproduces $A$ equally well, and the cost function provides no incentive to prefer $c \approx c_\text{ideal}$.
With $\mu = 10^{-6}$ (B), $c$ remains close to $c_\text{ideal}$, the success probability increases by almost three orders of magnitude, and the relative error is even slightly lower than for B$_0$.
Experiment C achieves lower overall error levels, while its success probability, $p_C \approx 0.3628$, remains almost identical to $p_B \approx 0.3597$ of experiment B.
}

%###########################################
%       References
%###########################################
\bibliography{references}
\end{document}